\documentclass[aip,jap,twocolumn,superscriptaddress,longbibliography,10pt]{revtex4-1}
\usepackage{amssymb}
\usepackage{amsmath}
\usepackage{bm}
\usepackage{glossaries}
\usepackage{fancyvrb}
\usepackage{inputenc}
\usepackage{fontenc}

\usepackage{algorithm}
\usepackage{algpseudocode}
\usepackage{silence} % stop revtex4-1 complaining
\usepackage{multirow}
\usepackage{makecell}
\usepackage{graphicx}
\usepackage[usenames,dvipsnames,svgnames]{xcolor}
\usepackage{hyperref}
\hypersetup{
pdfnewwindow=true,          % links in new window
colorlinks=true,            % false: boxed links; true: colored links
linkcolor=DarkBlue,         % color of internal links (change box color with linkbordercolor)
citecolor=DarkBlue,         % color of links to bibliography
filecolor=DarkBlue,         % color of file links
urlcolor=DarkBlue           % color of external links
}

\newacronym{aimd}{AIMD}{\textit{ab initio} molecular dynamics}
\newacronym{agf}{AGF}{atomistic Green function}
\newacronym{ald}{ALD}{anharmonic lattice dynamics}
\newacronym{ann}{ANN}{artificial neural network}
\newacronym{bte}{BTE}{Boltzmann transport equation}
\newacronym{dft}{DFT}{density-functional theory}
\newacronym{dp}{DP}{deep potential}
\newacronym{emd}{EMD}{equilibrium molecular dynamics}
\newacronym{gap}{GAP}{Gaussian approximation potential}
\newacronym{gpu}{GPU}{graphics processing unit}
\newacronym{hcacf}{HCACF}{heat current autocorrelation function}
\newacronym{bpnnp}{BPNNP}{Behler-Parrinello neural-network potential}
\newacronym{hnemd}{HNEMD}{homogeneous nonequilibrium molecular dynamics}
\newacronym{mtp}{MTP}{moment tensor potential}
\newacronym{ace}{ACE}{atomic cluster expansion}
\newacronym{md}{MD}{molecular dynamics}
\newacronym{mfp}{MFP}{mean free path}
\newacronym{ml}{ML}{machine learning}
\newacronym{mlp}{MLP}{machine-learned potential}
\newacronym{nemd}{NEMD}{nonequilibrium molecular dynamics}
\newacronym{nep}{NEP}{neuroevolution potential}
\newacronym{nn}{NN}{neural network}
\newacronym{rmse}{RMSE}{root-mean-square error}
\newacronym{snap}{SNAP}{spectral neighbor analysis potential}
\newacronym{snes}{SNES}{separable natural evolution strategy}
\newacronym{wte}{WTE}{Wigner transport equation}
\newacronym{zbl}{ZBL}{Ziegler-Biersack-Littmark}
\usepackage{siunitx}

\begin{document}

\title{Molecular dynamics simulations of heat transport using machine-learned potentials: A mini review and tutorial on GPUMD with neuroevolution potentials}

\author{Haikuan Dong}
\email{donghaikuan@163.com}
\affiliation{College of Physical Science and Technology, Bohai University, Jinzhou, P. R. China}

\author{Yongbo Shi}
%\email{syb461657001@126.com}
\affiliation{College of Physical Science and Technology, Bohai University, Jinzhou, P. R. China}

\author{Penghua Ying}
%\email{hityingph@163.com}
\affiliation{Department of Physical Chemistry, School of Chemistry, Tel Aviv University, Tel Aviv, 6997801, Israel}

\author{Ke Xu}
%\email{twtdq@stu.xmu.edu.cn}
\affiliation{Department of Electronic Engineering and Materials Science and Technology Research Center, The Chinese University of Hong Kong, Shatin, N.T., Hong Kong SAR, 999077, P. R. China}
 
\author{Ting Liang}
%\email{tingliang@link.cuhk.edu.hk}
\affiliation{Department of Electronic Engineering and Materials Science and Technology Research Center, The Chinese University of Hong Kong, Shatin, N.T., Hong Kong SAR, 999077, P. R. China}

\author{Yanzhou Wang}
%\email{yanzhou.wang@aalto.fi}
\affiliation{MSP group, QTF Centre of Excellence, Department of Applied Physics, Aalto University, FI-00076 Aalto, Espoo, Finland}

\author{Zezhu Zeng}
%\email{u3004964@connect.hku.hk}
\affiliation{The Institute of Science and Technology Austria, Am Campus 1, 3400 Klosterneuburg, Austria}

\author{Xin Wu}
%\email{xinwuchn97@gmail.com}
\affiliation{Department of Engineering Mechanics, School of Civil Engineering and Transportation, South China University of Technology, Guangzhou, Guangdong Province 510640, P. R. China}

\author{Wenjiang Zhou}
\affiliation{Department of Energy and Resources Engineering, Peking University, Beijing 100871, China}
\affiliation{School of Advanced Engineering, Great Bay University, Dongguan 523000, China}
%\email{wjzhou@stu.pku.edu.cn}

\author{Shiyun Xiong}
%email{syxiong@gdut.edu.cn}
\affiliation{Guangzhou Key Laboratory of Low-Dimensional Materials and Energy Storage Devices, School of Materials and Energy, Guangdong University of Technology,
Guangzhou 510006, P. R. China}

\author{Shunda Chen}
\email{phychensd@gmail.com}
\affiliation{Department of Civil and Environmental Engineering, George Washington University,
Washington, DC 20052, USA}

\author{Zheyong Fan}
\email{brucenju@gmail.com} 
\affiliation{College of Physical Science and Technology, Bohai University, Jinzhou, P. R. China}

\date{\today}

\begin{abstract}

Molecular dynamics (MD) simulations play an important role in understanding and engineering heat transport properties of complex materials. An essential requirement for reliably predicting heat transport properties is the use of accurate and efficient interatomic potentials. Recently, machine-learned potentials (MLPs) have shown great promise in providing the required accuracy for a broad range of materials. In this mini review and tutorial, we delve into the fundamentals of heat transport, explore pertinent MD simulation methods, and survey the applications of MLPs in MD simulations of heat transport. Furthermore, we provide a step-by-step tutorial on developing MLPs for highly efficient and predictive heat transport simulations, utilizing the neuroevolution potentials (NEPs) as implemented in the GPUMD package. Our aim with this mini review and tutorial is to empower researchers with valuable insights into cutting-edge methodologies that can significantly enhance the accuracy and efficiency of MD simulations for heat transport studies.
\end{abstract}

\maketitle
%\tableofcontents

\section{Introduction}

Heat transport properties are crucial for numerous applications \cite{volz2016epjp,li2012rmp}.
At the atomistic level, there are primarily three computational methods for heat transport \cite{gu2021jap}: \gls{md} simulations, methods related to \gls{bte}--including, more generally, quasi-harmonic Green-Kubo (QHGK) method \cite{isaeva_QHGK_NC_2019} and \gls{wte} approach \cite{Simoncelli-NatPhys2019,Simoncelli-PRX-2022}--combined with \gls{ald} (\gls{bte}-\gls{ald} for short), and \gls{agf}. Each method has its advantages and disadvantages \cite{gu2021jap}. This mini review and tutorial focuses on the \gls{md} methods. For the \gls{bte}-\gls{ald} and  \gls{agf} approaches, we refer interested readers to previous tutorials \cite{Ong2018jap,McGaughey2019jap,gu2021jap}. Our emphasis is on thermal conductivity, including finite systems, instead of thermal boundary conductance/resistance. For the latter, we suggest referring to a previous tutorial \cite{liang2018jap} and a review article\cite{chen2022rmp}.

Notable advantages distinguish \gls{md} from the other two methods. 
Firstly, \gls{md} can capture phonon-phonon scatterings at any order, while the other two methods are perturbative in nature and often consider only three-phonon scatterings (for \gls{bte}-\gls{ald}) or even completely ignore the anharmonicity (for \gls{agf}). 
Secondly, \gls{md} can naturally capture scatterings of phonons by other sources such as defects and mass disorder, extending its applicability to fluid systems that are beyond the reach of the other two methods. 
Thirdly, the computational cost of \gls{md} with classical potentials is usually linear with respect to the number of atoms, while it typically exhibits high-order polynomial scaling in the other two methods. 
Based on these considerations, \gls{md} proves particularly suitable for studying heat transport in strongly anharmonic or highly disordered systems.

Despite these advantages, \gls{md} simulations have grappled with challenges, particularly in terms of accuracy, over a considerable period of time. The predictive power of \gls{md} simulations is highly dependent on the accuracy of the classical potentials, which are mathematical models representing the potential energy surface of systems in terms of geometric information. The interatomic forces can be accurately computed using ab initio methods such as quantum-mechanical \gls{dft}, leading to the \gls{aimd} method, which has been applied to heat transport studies \cite{Marcolongo2016np,Carbogno2017prl,Kang2017prb,Knoop2023prb}.
A challenge in the \gls{aimd} approach is the high computational intensity, which imposes limitations on the size and timescales that can be effectively simulated.

\begin{figure}[htb]
\begin{center}
\includegraphics[width=\columnwidth]{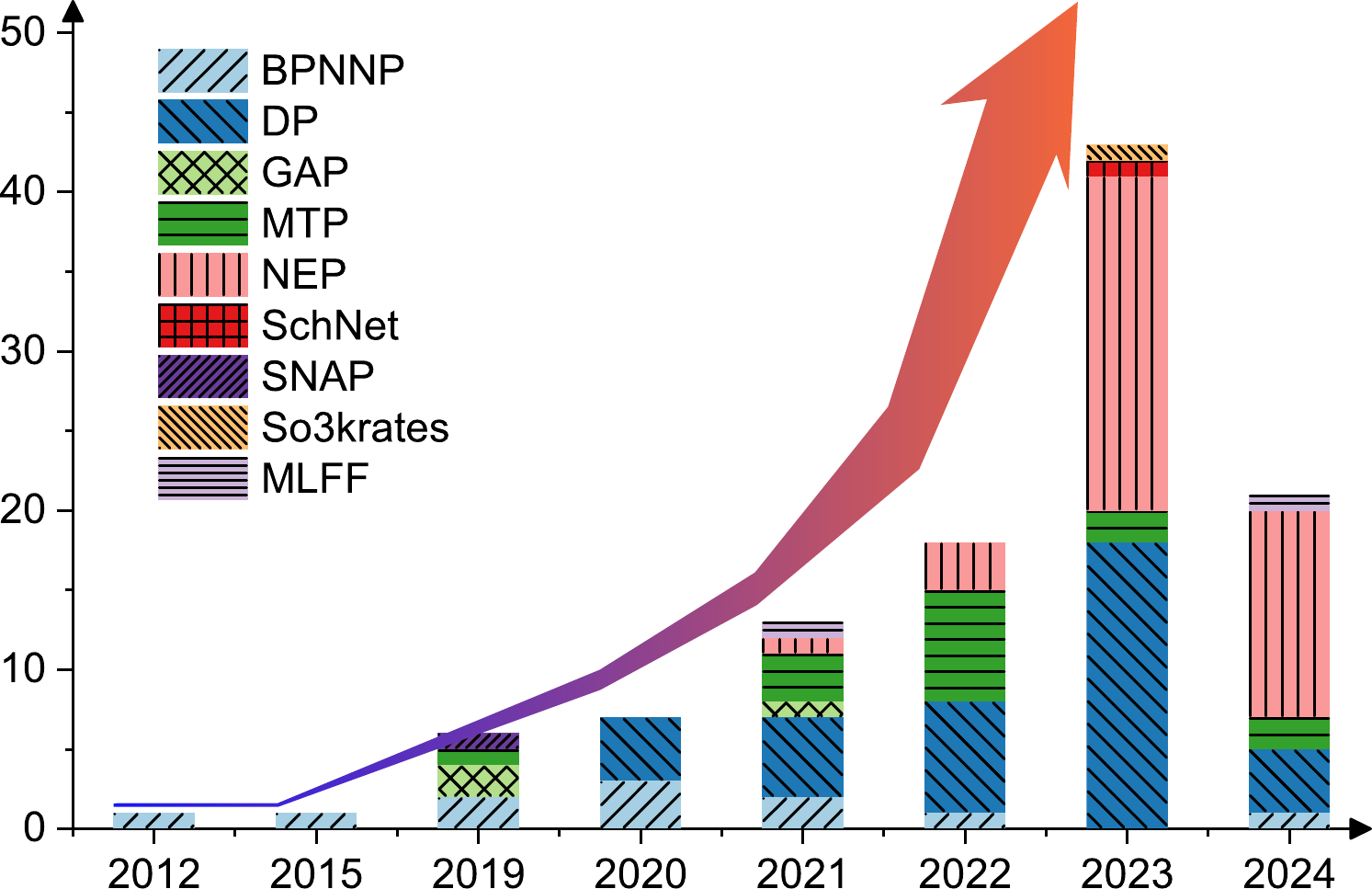}
\caption{Number of publications (up to March 10th, 2024) on heat transport \gls{md} simulations  using \glspl{mlp} as a function of year, with detailed information in Table~\ref{table:MLP-package} and  Table~\ref{table:mlp-kappa}. 
}
\label{figure:num_publications}
\end{center}
\end{figure}

Recently, a type of classical potentials based on \gls{ml} techniques, called \glspl{mlp}, has emerged as an effective framework for constructing highly accurate interatomic potentials.
Due to the flexible functional forms and a large number of fitting parameters in \glspl{mlp}, they can usually achieve significantly higher accuracy compared to traditional empirical potentials. Notable \gls{mlp} models, to name a few, include \gls{bpnnp} \cite{behler2007prl}, \gls{gap} and related \cite{Bartok2010prl,Caro2019prb,Byggmastar2022prm}, \gls{snap} \cite{Thompson2015jcp}, \gls{mtp} \cite{Novikov2021MLST}, \gls{dp} \cite{Wang2018cpc}, and \gls{ace} \cite{ACE-Drautz-PRB2019}. 
In this context, the recently developed \gls{nep} approach \cite{fan2021neuroevolution,fan2022jpcm,fan2022gpumd} simultaneously demonstrates excellent accuracy and outstanding computational efficiency, offering a distinctive advantage. Furthermore, \glspl{mlp} have been increasingly used in \gls{md} simulations, including heat transport simulations (see Fig.~\ref{figure:num_publications} for a general trend). 

Parallelization stands out as another key advancement in \gls{md} simulations, involving the deployment of parallel computing to take advantage of rapid hardware upgrades and speedups, where a large number of processors or cores work simultaneously to perform calculations, to augment computational efficiency and spatiotemporal scales of simulations. \textsc{gpumd} \cite{fan2017cpc}, short for  Graphics Processing Units Molecular Dynamics, represents a noteworthy development in this arena. \textsc{gpumd} is a versatile \gls{md} package fully implemented on \glspl{gpu}. This advancement facilitates the simulations of larger and more complex systems by leveraging the powerful parallel processing capabilities of \glspl{gpu}. For example, it has been demonstrated that \textsc{gpumd} can achieve a remarkable computational speed of \num{1.5e8} atom step/s (equivalent to a cost of $6.7\times 10^{-9}$ s/atom/step) in \gls{md} simulations using eight 80-gigabyte A100 graphics cards, enabling simulations up to 100 million atoms for high-entropy alloys employing a general-purpose unified \gls{nep} machine-learned potential for 16 elemental metals and their alloys \cite{song2023generalpurpose}.

In this mini review and tutorial, we dig into the fundamentals of heat transport, the relevant \gls{md} simulation methods, and the applications of \glspl{mlp} in \gls{md} simulations of heat transport. We use the \gls{nep} model \cite{fan2021neuroevolution,fan2022jpcm,fan2022gpumd} as implemented in the \textsc{gpumd} package \cite{fan2017cpc} to illustrate the various technical details involved. 
By completing this tutorial, the readers will gain both fundamental knowledge and practical skills to construct \glspl{mlp} and apply them in highly efficient and predictive \gls{md} simulations of heat transport.

\section{Fundamentals of heat transport and relevant MD simulation methods}

\subsection{Thermal conductivity and thermal conductance}

\subsubsection{Thermal conductivity}

Fourier's law describes the empirical relationship governing heat transport, expressed as:
\begin{equation}
Q_{\mu} = - \sum_{\nu}\kappa_{\mu\nu} \frac{\partial T}{\partial x_{\nu}}.
\end{equation}
Here $Q_{\mu}$ is the heat flux in the $\mu$ direction, $\frac{\partial T}{\partial x_{\nu}}$ is the temperature gradient in the $\nu$ direction, and $\kappa_{\mu\nu}$ is an element of the second-rank conductivity tensor \cite{nye1957book}. The heat flux measures the heat transport per unit time and per unit area, typically measured in W m$^{-2}$. The thermal conductivity is commonly expressed in units of W m$^{-1}$ K$^{-1}$.

When the coordinate axes align with the principal axes of the conductivity tensor, thermal transport decouples in different directions, yielding a diagonal thermal conductivity tensor with three nonzero elements: $\kappa_{xx}$, $\kappa_{yy}$, and $\kappa_{zz}$. These are commonly denoted as $\kappa_{x}$, $\kappa_{y}$, and $\kappa_{z}$ for simplicity.
For isotropic 3D systems, we usually define a conductivity scalar $\kappa$ in terms of the trace of the tensor:
$
\kappa = (\kappa_{x} + \kappa_{y} + \kappa_{z})/3
$.
For isotropic 2D systems, we usually define a conductivity scalar for the planar components:
$
\kappa = (\kappa_{x} + \kappa_{y})/2
$.
For quasi-1D systems, it is only meaningful to define the conductivity in a single direction. For simplicity, from here on we work with the conductivity scalar $\kappa$ unless it is necessary to consider the conductivity tensor. 

\subsubsection{Thermal conductance}

In macroscopic transport (the meaning of which will become clear soon), thermal conductance $K$ is related to thermal conductivity by:
\begin{equation}
K = \kappa \frac{A}{L},
\label{equation:K_and_kappa}
\end{equation}
where $A$ is the cross-sectional area and $L$ is the system length along the transport direction. This relation is similar to that between electrical conductance and electrical conductivity one learns in high school. Usually, thermal conductivity is considered an intrinsic property of a material, and  thermal conductance depends on the geometry ($A$ and $L$). However, complexities emerge when examining heat transport at the nanoscale or mesoscale.

At the nanoscale, the conventional concept of conductivity may lose its validity \cite{datta1995}. For example, thermal transport in materials with high thermal conductivity, such as diamond at the nanoscale, is almost \textit{ballistic}, meaning the conductance changes little with increasing system length $L$. In this case, if we assume that Eq. (\ref{equation:K_and_kappa}) still holds, then the thermal conductivity $\kappa$ cannot be regarded as a constant but as a function of the system length, $\kappa=\kappa(L)$. This deviates from the conventional (macroscopic) concept of thermal conductivity.

Rather than adhering strictly to Eq. (\ref{equation:K_and_kappa}), one can generalize the relation between conductance and conductivity as follows:
\begin{equation}
\frac{1}{K} = \frac{1}{K_0} + \frac{1}{\kappa} \frac{L}{A},
\label{equation:K_and_kappa_diff}
\end{equation}
where $K_0$ is the ballistic thermal conductance of the material. 
The term $\kappa$ in Eq. (\ref{equation:K_and_kappa_diff}) refers to the \textit{diffusive thermal conductivity}, the conventional thermal conductivity defined in the macroscopic limit ($L \to \infty$) where the phonon transport is diffusive. By contrast, the length-dependent thermal conductivity $\kappa(L)$ defined in Eq. (\ref{equation:K_and_kappa}) is usually called the \textit{apparent thermal conductivity} or \textit{effective thermal conductivity}. In the diffusive limit, the apparent thermal conductivity $\kappa(L)$ defined in Eq. (\ref{equation:K_and_kappa}) approaches the diffusive conductivity $\kappa$ defined in Eq. (\ref{equation:K_and_kappa_diff}), as expected. 

\begin{figure}[htb]
\begin{center}
\includegraphics[width=\columnwidth]{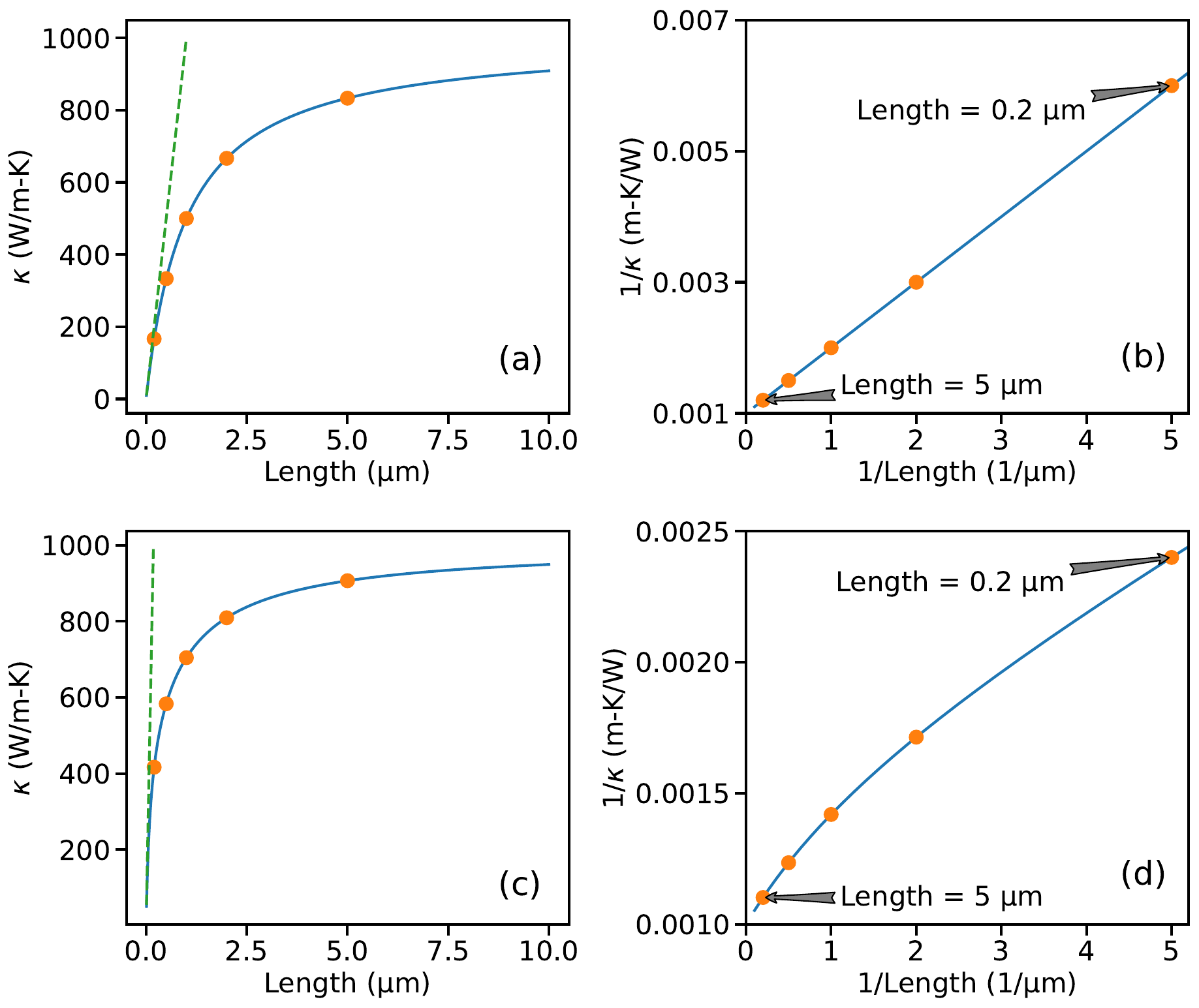}
\caption{Ballistic-to-diffusive transition of the apparent thermal conductivity $\kappa(L)$. (a)-(b) a toy model with a single phonon MFP of 1 $\mu$m and a diffusive thermal conductivity of $\kappa=1000$ W/mK; (c)-(d) a toy model with two phonon MFPs, one of $0.1$ $\mu$m, the other $1$ $\mu$m, with diffusive conductivity of 500 W/mK. The dots in each panel represent a few special lengths, from 0.2 $\mu$m to 5 $\mu$m. In (a) and (c), the dashed lines represent the ballistic limit.}
\label{figure:ballistic_to_diffusive}
\end{center}
\end{figure}

By comparing Eq. (\ref{equation:K_and_kappa}) and Eq. (\ref{equation:K_and_kappa_diff}), we obtain the following relation between the apparent thermal conductivity $\kappa(L)$ and the diffusive thermal conductivity $\kappa$:
\begin{equation}
\frac{1}{\kappa(L)} \frac{L}{A} = \frac{1}{K_0} + \frac{1}{\kappa} \frac{L}{A}.
\label{equation:kappa_and_kappa_diff}
\end{equation}
From this, we have
\begin{equation}
\frac{1}{\kappa(L)} = \frac{1}{\kappa} 
\left( 
1+ \frac{\kappa A}{K_0 L}
\right).
\label{equation:ballistic_to_diffusive}
\end{equation}
It is more common to use thermal conductance per unit area $G$, which is
defined as
\begin{equation}
G \equiv \frac{K}{A}.
\end{equation}
The corresponding ballistic conductance per unit area is $G_0=K_0/A$. Using this, we have 
\begin{equation}
\frac{1}{\kappa(L)} = \frac{1}{\kappa} 
\left( 
1+ \frac{\kappa /G_0}{L}
\right).
\end{equation}
The ratio between the diffusive conductivity and the ballistic conductance per unit area defines a phonon \gls{mfp}:
\begin{equation}
\lambda \equiv \frac{\kappa}{ G_0}.
\label{equation:mfp_from_G}
\end{equation}
In terms of the phonon MFP, we have 
\begin{equation}
\frac{1}{\kappa(L)} = \frac{1}{\kappa} 
\left( 
1+ \frac{\lambda}{L}
\right).
\label{equation:ballistic_to_diffusive_lambda}
\end{equation}
This is known as the ballistic-to-diffusive transition formula for the length dependent thermal conductivity. Figure \ref{figure:ballistic_to_diffusive} schematically shows the ballistic-to-diffusive transition behavior. 

The above discussion is simplified in the sense that no channel dependence of the thermal transport has been taken into account. Different channels usually have different \glspl{mfp} and diffusive conductivities. In general, both the conductivity and the MFP are frequency dependent and we can generalize Eq. (\ref{equation:ballistic_to_diffusive_lambda}) to
\begin{equation}
\frac{1}{\kappa(\omega,L)} = \frac{1}{\kappa(\omega)} 
\left( 
1+ \frac{\lambda(\omega)}{L}
\right).
\label{equation:ballistic_to_diffusive_lambda_omega}
\end{equation}
With $\kappa(\omega,L)$, we can obtain the apparent thermal conductivity at \textit{any} length $L$ as 
\begin{equation}
\label{equation:kappa_L_from_integration}
    \kappa(L) = \int_0^{\infty} \frac{d\omega}{2\pi}\kappa(\omega,L).
\end{equation}

We use two toy models to illustrate the above-discussed concepts. In the first model, we assume that there is only one phonon \gls{mfp} of 1 $\mu$m and a diffusive thermal conductivity of $\kappa=1000$ W m$^{-1}$ K$^{-1}$. Then the ballistic conductance is $\kappa/\lambda=1$ GW m$^{-2}$ K$^{-1}$. Then the apparent thermal conductivity $\kappa(L)$ is given by Eq. (\ref{equation:ballistic_to_diffusive_lambda}), as shown in Figs.~\ref{figure:ballistic_to_diffusive}(a) and \ref{figure:ballistic_to_diffusive}(b). In this case, $1/\kappa(L)$ varies linearly with $1/L$. In the second model, we assume that there are two phonon modes, one with a \gls{mfp} of $0.1$ $\mu$m, and the other $1$ $\mu$m, both having a diffusive conductivity of 500 W m$^{-1}$ K$^{-1}$. Then the ballistic conductances for these two modes are 5 GW m$^{-2}$ K$^{-1}$ and 0.5 GW m$^{-2}$ K$^{-1}$, respectively. The higher ballistic conductance in the second toy model can be visualized in Fig. \ref{figure:ballistic_to_diffusive}(c). Although the apparent thermal conductivity for each mode follows Eq. (\ref{equation:ballistic_to_diffusive_lambda}), when combined, $1/\kappa(L)$ does \textit{not} exhibit linearity with $1/L$. This is an important feature for realistic materials with a general \gls{mfp} spectrum $\kappa(\omega)$.

\subsection{Heat flux and heat current}

The heat flux is defined as the time derivative of the sum of the moments of the site energies of the particles in the system \cite{mcquarrie2000book}:
\begin{equation}
  \bm{Q} \equiv \frac{1}{V} \frac{d}{d t} \sum_i \bm{r}_i E_i.
\end{equation}
The site energy $E_i$ is the sum of the kinetic energy $m_i \bm{v}_i^2/2$ and the potential energy $U_i$.
Here $m_i$, $\bm{r}_i$, and $\bm{v}_i$ are the mass, position, and velocity of particle $i$, respectively, and $V$ is the controlling volume for the particles, which is usually the volume of the simulation box, but can also be specifically defined for low-dimensional systems simulated with vacuum layers. In \gls{md} simulations, it is usually more convenient to work on the heat current that is an extensive quantity: 
\begin{equation}
\bm{J} \equiv V \bm{Q} 
\end{equation}
It is clear that the total heat current can be written as two terms:
\begin{equation}
\bm{J} =  \bm{J}_{\textmd{kin}} + \bm{J}_{\textmd{pot}},
\end{equation}
where the first term is the kinetic or convective part,
\begin{equation}
\bm{J}_{\textmd{kin}} =  \sum_i \bm{v}_i E_i
\end{equation}
and the second term is called the potential part,
\begin{equation}
  \bm{J}_{\textmd{pot}} 
  = \sum_i \bm{r}_i (\bm{F}_i \cdot \bm{v}_i)
  + \sum_i \bm{r}_i \frac{d U_{i}}{d t}.
  \label{equation:j_pot}
\end{equation}
The expression above involves absolute positions and is thus not directly applicable to periodic systems.
To derive an expression that can be used for periodic systems, we need to discuss potential energy and interatomic force first.

\begin{figure}[htb]
\begin{center}
\includegraphics[width=\columnwidth]{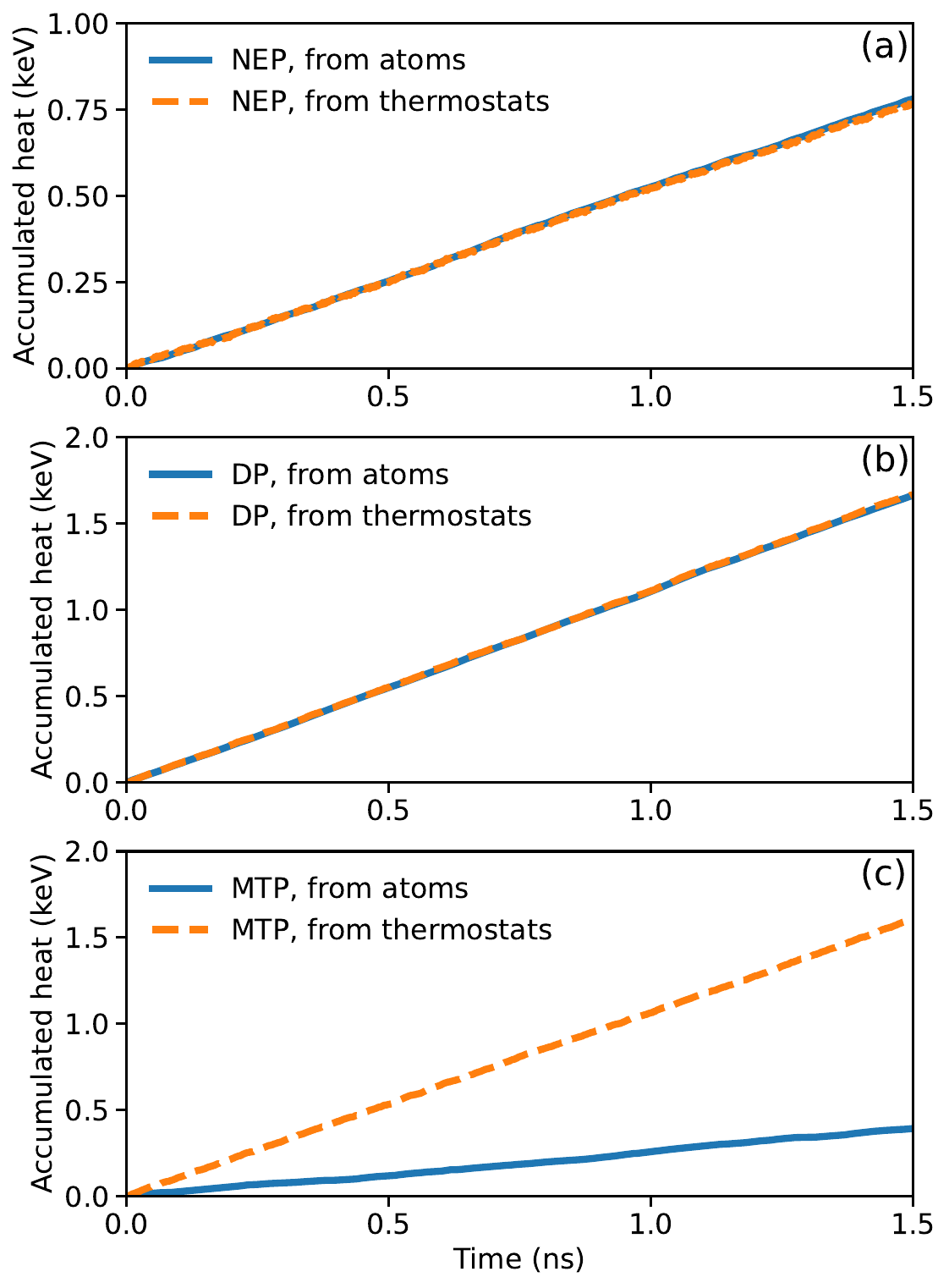}
\caption{Accumulated heat as a function of time in non-equilibrium steady state simulated with (a) \gls{nep}, (b) \gls{dp}, and (c) \gls{mtp}, using \textsc{gpumd} \cite{fan2017cpc} (for \gls{nep}) or \textsc{lammps} \cite{Thompson2022cpc} (for \gls{dp} and \gls{mtp}). }
\label{figure:heatflux}
\end{center}
\end{figure}

For the \glspl{mlp} discussed in this tutorial, the total potential energy $U$ of a system can be written as the sum of site potentials $U_i$:
\begin{equation}
\label{equation:U}
U=\sum_{i=1}^N U_i.
\end{equation}
The site potential can have different forms in different potential models. 
A well-defined force expression for general many-body potentials that explicitly respects Newton's third law  has been derived as \cite{fan2015prb}:
\begin{equation}
\label{equation:F_i}
  \bm{F}_{i} = \sum_{j \neq i} \bm{F}_{ij},
\end{equation}
where
\begin{equation}
\label{equation:F_ij}
\bm{F}_{ij} = - \bm{F}_{ji} =
  \frac{\partial U_{i}}{\partial \bm{r}_{ij}} -
  \frac{\partial U_{j}}{\partial \bm{r}_{ji}}.
\end{equation}
Here, $\partial U_{i}/\partial \bm{r}_{ij}$ is a shorthand notation for a vector with Cartesian components $\partial U_{i}/\partial x_{ij}$, $\partial U_{i}/\partial y_{ij}$, and $\partial U_{i}/\partial z_{ij}$. The atomic position difference is defined as
\begin{equation}
    \bm{r}_{ij} \equiv \bm{r}_{j} - \bm{r}_{i}.
\end{equation}
Using the force expression, the heat current can be derived to be \cite{fan2015prb}:
\begin{equation}
\bm{J}_{\textmd{pot}} = \sum_i \sum_{j \neq i} \bm{r}_{ij}
\left(
      \frac{\partial U_j}{\partial \bm{r}_{ji}} \cdot \bm{v}_i
\right).
\label{equation:j_pot_3}
\end{equation} 
From the definition of virial tensor 
\begin{equation}
\textbf{W} = \sum_i \textbf{W}_i = \sum_i \bm{r}_{i} \otimes \bm{F}_i
\end{equation}
and the force expression Eq.~(\ref{equation:F_i}), we have 
\begin{equation}
 \textbf{W} 
 = -\frac{1}{2} \sum_i \sum_{j \neq i}\bm{r}_{ij} \otimes \bm{F}_{ij}.
\end{equation}
Using the explicit force expression Eq.~(\ref{equation:F_ij}), we can also express the per-atom virial as 
\begin{equation}
 \label{equation:per_atom_virial}
 \textbf{W}_i = \sum_{j \neq i}\bm{r}_{ij} \otimes \frac{\partial U_j}{\partial \bm{r}_{ji}} .
\end{equation}
Therefore, the heat current can be neatly written as:
\begin{equation}
\label{equation:j_pot_pair_stress}
  \bm{J}_{\textmd{pot}} = \sum_{i} \textbf{W}_{i} \cdot \bm{v}_i.
\end{equation}
This expression, which involves relative atom positions only, is applicable to periodic systems and has been implemented in the \textsc{gpumd} package \cite{fan2017cpc} for all the supported interatomic potentials, including \gls{nep}. 
The current implementation of the heat current in \textsc{lammps} \cite{Thompson2022cpc} is generally incorrect for many-body potentials, and corrections to \textsc{lammps} have only been done for special force fields
\cite{boone2019jctc,surblys2019pre}. For any \gls{mlp} that interfaces with \textsc{lammps}, one must use the full 9 components of the per-atom virial and provide a correct implementation of Eq. (\ref{equation:per_atom_virial}). \gls{nep} has an interface for \textsc{lammps} that meets this requirement. To the best of our knowledge, among the other publicly available \gls{mlp} packages,  only \textsc{deepmd}  \cite{Wang2018cpc} (after the work of Tisi \textit{et al.} \cite{Tisi2021PRB}) and \textsc{aenet} \cite{Nongnuch2016cms} (after the work of Shimamura \textit{et al.} \cite{Shimamura2020jcp}) have implemented the heat current correctly. The heat current is also correctly formulated \cite{tisi2024thermal} for a \gls{mlp} based on the smooth overlap of atomic positions \cite{bartok2013prb}. Contrarily, the widely used \gls{mtp} method \cite{Novikov2021MLST} (as implemented in Ref. \onlinecite{mlip}), for example, exhibits an incorrect implementation of the heat current, as demonstrated in Fig.~\ref{figure:heatflux}. According to energy conservation, the accumulated heat from the atoms [cf. Eq.~(\ref{equation:j_pot_pair_stress})] should match that from the thermostats [cf. Eq.~(\ref{equation:Q_thermostat})], allowing for only small fluctuations. It is evident that both \gls{dp} and \gls{nep} exhibit this property, whereas \gls{mtp} does not. Details on the calculations are provided in Appendix~\ref{section:appendix}.

Note that the above formulation of heat current has been derived specifically for \textit{local} \glspl{mlp} with atom-centered descriptors. For \textit{semilocal} message-passing-based \glspl{mlp}, the formulation of heat current has been shown by Langer \textit{et al.} \cite{Langer2023PRB,Langer2023jcp} to be more complicated.

\subsection{Overview of MD-based methods for heat transport}

In the following, we review the heat transport \gls{md} methods implemented in the \textsc{gpumd} package, including \gls{emd}, \gls{nemd}, \gls{hnemd}, and spectral decomposition. While the approach-to-equilibrium method \cite{Daly2002pb,Daly2002prb,Lampin2013jap} can in principle be realized in \textsc{gpumd}, our discussion will primarily focus on the other three methods that have been widely employed with \textsc{gpumd}.

\subsubsection{The EMD method \label{section:emd}}

The \gls{emd} method is based on the Green-Kubo relation for thermal transport\cite{kubo1957jpsj}:
\begin{equation}
\label{equation:kappa_t_emd}
\kappa_{\mu\nu}(t) = \frac{1}{k_{\rm B} T^2V} \int_0^{t} dt' C_{\mu\nu} (t'),
\end{equation}
where $C_{\mu\nu}(t)$ is the \gls{hcacf}
\begin{equation}
 C_{\mu\nu}(t) = \langle J_{\mu}(0) J_{\nu}(t) \rangle_{\rm e}.
\end{equation}
The equations above define the running thermal conductivity, which is a function of the \textit{correlation time} $t$. In MD simulations, the correlation function is defined as
\begin{equation}
\langle J_{\mu}(0) J_{\nu}(t) \rangle_{\rm e} \approx
\frac{1}{t_{\rm p}} \int_0^{t_{\rm p}}  J_{\mu}(\tau) J_{\nu}(t+\tau) d\tau.
\end{equation}
where $t_{\rm p}$ is the \textit{production time} within which the heat current data are sampled. This production run should be in an equilibrium ensemble (as indicated by the subscript ``e'' in the \gls{hcacf} expression), usually $NVE$, but $NVT$ with a \textit{global} thermostat can also be used.
Thermal conductivity in the diffusive limit is obtained by taking the limit of $t\to \infty$, but in practice, this limit can be well approximated at an appropriate $t$. One also needs to ensure that the simulation cell is sufficiently large to eliminate finite-size effects \cite{ladd1986prb,Chen-finite-size-PRE2014,wang2017jap}. 

\subsubsection{The NEMD method \label{section:nemd}}

The \gls{nemd} method is a nonequilibrum and inhomogeneous method that involves implementing a pair of heat source and sink using a thermostatting method or equivalent.
There are two common relative positions of the source and sink in the \gls{nemd} method, corresponding to two typical simulation setups. In one setup, the source and sink are separated by half of the simulation cell length $L$, and periodic boundary conditions are applied along the transport direction. Heat flows from the source to the sink in two opposite directions in this periodic boundary setup. In the other setup, the source and sink separated by $L$ are located at the two ends of the system. Fixed boundary conditions are applied along the transport direction to prevent sublimation of the atoms in the heat source and sink. Heat flows from the source to the sink in one direction in this fixed boundary setup. It has been established \cite{xu2018msmse} that the effective length in the periodic boundary setup is only $L/2$. This factor must be taken into account when comparing results from the two setups. 

When the system reaches a steady state, a temperature profile with a definite temperature gradient $\nabla T$ will be established. Meanwhile, a steady heat flux $Q$ will be generated. With these, one can obtain the apparent thermal conductivity $\kappa(L)$ of a system of finite length $L$ according to Fourier's law,
\begin{equation}
\label{equation:kappa_L_NEMD}
\kappa(L) =
\frac{Q}{|\nabla T|},
\end{equation}
in the linear response regime where the temperature gradient $|\nabla T|$ across the system is sufficiently small.
It has been observed that the local Langevin thermostat outperforms the global Nos\'{e}-Hoover thermostat \cite{nose1984jcp,hoover1985pra} in generating temperature gradients \cite{li2019jcp}. It has also been demonstrated that the temperature gradient should be directly calculated  from the temperature difference $|\nabla T| =\Delta T/L$ rather than through fitting part of the temperature profile \cite{li2019jcp}. This is to ensure that the contact resistance is also included, and the total thermal conductance is given by
\begin{equation}
    G(L) = \frac{Q}{|\Delta T|}.
\end{equation}

The steady-state heat flux can be computed either microscopically or from the energy exchange rate $dE/dt$ in the thermostatted regions and cross-sectional area $A$ as 
\begin{equation}
\label{equation:Q_thermostat}
    Q=\frac{1}{A}\frac{dE}{dt},
\end{equation}
based on energy conservation. The two approaches must generate the same result, and they have been used to validate the implementation of heat flux in several \glspl{mlp}, as shown in Fig.~\ref{figure:heatflux}.

A common practice in using the \gls{nemd} method is to extrapolate to the limit of infinite length based on the results for a few finite lengths. It is important to note that linear extrapolation is usually insufficient, as suggested even by the toy-model results shown in Fig.~\ref{figure:ballistic_to_diffusive}(d).

\subsubsection{The HNEMD method \label{section:hnemd}}

In the \gls{hnemd} method, an external force of the form \cite{fan2019prb}
\begin{equation}
\bm{F}_{i}^{\rm ext} = E_i \bm{F}_{\rm e} +  \bm{F}_{\rm e} \cdot  \mathbf{W}_i
\end{equation}
is added to each atom to drive the system out of equilibrium, inducing a nonequilibrum heat current (note the subscript ``ne''):
\begin{equation}
\langle \bm{J}(t)\rangle_{\rm ne}
=
\left(
\frac{1}{k_{\rm B}T}\int_0^tdt'\langle \bm{J}(0)\otimes \bm{J}(t')\rangle_{\rm e}
\right)
\cdot \bm{F}_{\rm e}.
\label{equation:J(t)}
\end{equation}
The driving force parameter $\bm{F}_{\rm e}$ is of the dimension of inverse length.
The quantity in the parentheses is proportional to the running thermal conductivity tensor and we have 
\begin{equation}
\label{equation:kappa_t_hnemd}
\frac{\langle J^{\mu}_{\rm q}(t)\rangle_{\rm ne}}{TV}
= \sum_{\nu} \kappa^{\mu\nu}(t) F_{\rm e}^{\nu}.
\end{equation}
This provides a way of computing the thermal conductivity.
In the \gls{hnemd} method, the system is in a homogeneous nonequilibrium state because there is no explicit heat source and sink. The system is periodic in the transport direction and heat flows circularly under the driving force. Because of the absence of heat source and sink, no boundary scattering occurs for the phonons and the \gls{hnemd} method is similar to the \gls{emd} method in terms of finite-size effects. 

\subsubsection{Spectral decomposition}

In the framework of the \gls{nemd} and \gls{hnemd} methods, one can also calculate spectrally decomposed thermal conductivity (or conductance) using the virial-velocity correlation function \cite{fan2017prb,fan2019prb}
\begin{equation}
    \bm{K}(t) = \left\langle \sum_i \mathbf{W}_i(0) \cdot \bm{v}_i(t) \right\rangle_{\rm ne}.
\end{equation}
In terms of this, the thermal conductance in \gls{nemd} simulation can be decomposed as follows: 
\begin{equation}
G = \int_{-\infty}^{+\infty} \frac{d\omega}{2\pi} G(\omega);
\end{equation}
\begin{equation}
\label{equation:G_omega}
G(\omega) = \frac{2}{V\Delta T} \int_{-\infty}^{+\infty} e^{i\omega t} K^{\mu}(t) dt. 
\end{equation}
The thermal conductivity in \gls{hnemd} simulation can be decomposed as follows: 
\begin{equation}
\kappa_{\mu\nu} = \int_{-\infty}^{+\infty} \frac{d\omega}{2\pi} \kappa_{\mu\nu}(\omega);
\end{equation}
\begin{equation}
\label{equation:kappa_omega}
\frac{2}{VT} \int_{-\infty}^{+\infty} e^{i\omega t} K^{\mu}(t) dt= \sum_{\nu} \kappa_{\mu\nu}(\omega){F_{\rm e}}^\nu.
\end{equation}
The virial-velocity correlation function here is essentially the force-velocity correlation function defined for a (physical or imaginary) interface \cite{saaskilahti2014prb,saaskilahti2015prb}.

The spectral quantities allow for a feasible quantum-statistical correction \cite{gu2021jap,Lv2016njp} for strongly disordered systems where phonon-phonon scatterings are not dominant. For example, the spectral thermal conductivity can be quantum-corrected by multiplying the factor
\begin{equation}
\label{equation:quantum-correction-factor}
\frac{x^2e^x}{(e^x-1)^2},
\end{equation}
where $x=\hbar\omega/k_{\rm B} T$.

There are other spectral/modal analysis method implemented in \textsc{gpumd}, such as the Green-Kubo modal analysis method \cite{Gabourie2021prb} and the Homogeneous non-equilibrium modal analysis method \cite{Lv2016njp}, but we will not demonstrate their usage in this tutorial.

\section{Review of MD simulation of heat transport using MLPs\label{section:MD-methods}}

Several \glspl{mlp} have been used for heat transport with \gls{md} simulations, including
\gls{bpnnp} \cite{behler2007prl},
\gls{gap} \cite{Bartok2010prl},
\gls{snap} \cite{Thompson2015jcp},
\gls{mtp} \cite{Novikov2021MLST},
\gls{dp} \cite{Wang2018cpc},
MLFF \cite{Jinnouchi2019prl},
\gls{nep} \cite{fan2021neuroevolution},
SchNet \cite{Schnet2017},
and
So3krates \cite{Frank2022so3krates}.
Table~\ref{table:MLP-package} lists the relevant \gls{mlp} packages implementing these \glspl{mlp}. 

\begin{table*}[ht!]
\caption{The \glspl{mlp} and their implementation packages that have been used in \gls{md} simulations of heat transport.}
\begin{center}
\begin{tabular}{ l l l l l}
\hline
Year & MLP & Package   & Code repository or official website  \\
\hline
2007 & \gls{bpnnp} & \textsc{runner}   &  \url{https://theochemgoettingen.gitlab.io/RuNNer}  \\
 &  & \textsc{aenet}    & \url{https://github.com/atomisticnet/aenet}  \\
 &  & \textsc{kliff}    & \url{https://github.com/openkim/kliff} \\
2010 & \gls{gap}   &\textsc{quip}    &  \url{https://github.com/libAtoms/QUIP}  \\
2015 & \gls{snap}  &\textsc{fitsnap} &  \url{https://github.com/FitSNAP/FitSNAP} \\
2016 & \gls{mtp}   &\textsc{mlip}    &  \url{https://gitlab.com/ashapeev/mlip-2}  \\
2017 & SchNet      & \textsc{schnetpack}  & \url{https://github.com/atomistic-machine-learning/schnetpack}  \\
2018 & \gls{dp}    &\textsc{deepmd-kit}  &  \url{https://github.com/deepmodeling/deepmd-kit}\\
2019 & MLFF  &\textsc{vasp}  &  \url{https://www.vasp.at}\\
2021 & \gls{nep}   &\textsc{gpumd}   &  \url{https://github.com/brucefan1983/GPUMD} \\
2024 & So3krates   &\textsc{mlff}    &  \url{https://github.com/thorben-frank/mlff}  \\
\hline
\end{tabular}
\end{center}
\label{table:MLP-package}
\end{table*}

The pioneering \gls{bpnnp} model, developed by Behler and Parrinello \cite{behler2007prl}, has been implemented in various packages, including \textsc{runner} \cite{behler2007prl}, \textsc{aenet} \cite{Nongnuch2016cms}, and \textsc{kliff} \cite{wen2022kliff}. The \gls{dp}, MLFF, \gls{nep}, \gls{gap}, \gls{mtp}, \gls{snap}, SchNet, and So3krates models are implemented in \textsc{deepmd-kit}, \textsc{vasp}, \textsc{gpumd}, \textsc{quip}, \textsc{mlip}, \textsc{fitsnap}, \textsc{schnetpack}, and \textsc{mlff} respectively.

Most \gls{mlp} packages are interfaced to \textsc{lammps} \cite{Thompson2022cpc} to perform \gls{md} simulations, while \gls{nep} is native to \textsc{gpumd} \cite{fan2017cpc} but can also be interfaced to \textsc{lammps}. The MLFF method implemented in \textsc{vasp} is an on-the-fly \gls{mlp} that integrates seamlessly into \gls{aimd} simulations.

Table~\ref{table:mlp-kappa} compiles the publications up to today that have used \gls{md} simulations driven by \glspl{mlp} for thermal transport studies. Note that our focus is on studies using \gls{md} simulations, excluding those solely based on the \gls{bte}-\gls{ald} approach. The number of publications up to March 10th, 2024 for each \gls{mlp} is shown in Fig.~\ref{figure:num_publications}. 

\begin{table*}[!]
\caption{Applications of \glspl{mlp} in MD simulations of heat transport up to March 10th, 2024. %Only the first author of each paper is listed here. More information about the \glspl{mlp} here can be found in Table~\ref{table:MLP-package}.
}
\begin{center}
\begin{tabular}{ l  l l l l l l l}
\hline
\hline
\gls{mlp} & Year & Reference   & Material(s)  & Year & Reference   & Material(s)\\
\hline  

%%%%%%%%%%%%%%%%%%%%%%%%%%%%%%%%% bpnnp
\hline
\gls{bpnnp} 
 & 2012 & Sosso \cite{Sosso2012prb}  & Amorphous GeTe 
 & 2015 & Campi  \cite{Campi2015jap}  & GeTe \\
 & 2019 & Bosoni  \cite{Bosoni2019jpd} &  GeTe nanowires 
 & 2019 & Wen  \cite{wen2019prb}   & 2D graphene \\
 & 2020 & Cheng  \cite{Cheng2020PRL}  & Liquid hydrogen 
 & 2020 & Mangold  \cite{Mangold2020jap}  & Mn$_x$Ge$_y$ \\
 & 2020 & Shimamura  \cite{Shimamura2020jcp} &  Ag$_2$Se  
 & 2021 & Han  \cite{Han2021cms} &  Sn \\
 & 2021 & Shimamura  \cite{Shimamura2021cpl}  &  Ag$_2$Se 
 & 2022 & Takeshita  \cite{Takeshita2022jpcs} & Ag$_2$Se \\
 & 2024 & Shimamura  \cite{Shimamura2024cpc} &  Ag$_2$Se \\

%%%%%%%%%%%%%%%%%%%%%%%%%%%%%%%%% GAP
 \hline 
\gls{gap} & 2019 & Qian  \cite{qian2019mtp} & Silicon  
&2019 & Zhang  \cite{zhang2019jap}  & 2D silicene \\ 
&2021 & Zeng  \cite{zeng2021prb} &  Tl$_3$VSe$_4$ \\ 

%%%%%%%%%%%%%%%%%%%%%%%%%%%%%%%%% SNAP
\hline
\gls{snap} & 2019 & Gu  \cite{gu2019thermalCMS}   & MoSSe alloy\\ 

%%%%%%%%%%%%%%%%%%%%%%%%%%%%%%%%% MTP
\hline
\gls{mtp} & 2019 & Korotaev  \cite{korotaev2019prb} &   CoSb$_3$  
& 2021 & Liu  \cite{liu2021jpcm} & SnSe \\ 
& 2021 & Yang  \cite{Yang2021prb} & CoSb$_3$ and Mg$_3$Sb$_2$ 
& 2021 & Zeng  \cite{zeng2021mtp}   & BaAg$_2$Te$_2$ \\ 
& 2022 & Attarian  \cite{attarian2022thermophysical}  & FLiBe 
& 2022 & Ouyang \cite{Ouyang2022ijhmt} & SnS \\
& 2022 & Ouyang  \cite{quyang2022prb} &  BAs and Diamond  
& 2022 & Mortazavi  \cite{mortazavi2022fc,mortazavi2022carbon,mortazavi2022coat}   & Graphyne, 2D BCN, C$_{60}$\\
& 2022 & Sun  \cite{sun2022afm} & Bi$_2$O$_2$Se
& 2023 & Mortazavi  \cite{mortazavi2023carbon,mortazavi2023mtn,mortazavi2023hexagonal}  & C$_{60}$, C$_{36}$ and B$_{40}$ networks \\
& 2023 & Wang  \cite{wang2023b} & Cs$_2$AgPdCl$_5$ etc. 
& 2023 & Zhu  \cite{Zhu2023PRB} & CuSe   \\
& 2024 & Chang \cite{Chang2024msmse} &  PbSnTeSe and PbSnTeS
& 2024 & Wieser  \cite{wieser2024machine}   & MOF crystals \\
% & Preprint & Zhou \cite{zhou2024flattening} & La$_2$Zr$_2$O$_7$ \\

%%%%%%%%%%%%%%%%%%%%%%%%%%%%%%%%% SCHNET
\hline
SchNet & 2023 & Langer  \cite{Langer2023PRB}   & ZrO$_2$\\ 

%%%%%%%%%%%%%%%%%%%%%%%%%%%%%%%%% DP
\hline 
\gls{dp} & 2020 & Dai  \cite{dai2020theoretical} & ZrHfTiNbTaC alloy 
&2020 & Li  \cite{li2020apl} &  $\beta$-Ga$_2$O$_3$ \\ 
&2020 & Li  \cite{li2020unified} & Silicon 
&2020 & Pan  \cite{pan2020dft}  & ZnCl$_2$ \\
&2021 & Bu  \cite{bu2021local}  & KCl-CaCl$_2$ molten salt  
&2021 & Dai  \cite{dai2021temperature}  & TiZrHfNbTaB alloy  \\
&2021 & Deng  \cite{deng2021thermal} &  MgSiO$_3$ liquid 
&2021 & Liu  \cite{liu2021thermal}   & Al  \\
&2021 & Tisi  \cite{Tisi2021PRB} &  Water 
&2022 & Gputa  \cite{gupta2022strongly}  & Cu$_7$PSe$_6$  \\
&2022 & Huang  \cite{huang2022nanotwinning}  & B$_{12}$P$_2$ 
&2022 & Liang  \cite{liang2022machine} & MgCl$_2$-NaCl eutectic  \\
&2022 & Pegolo  \cite{pegolo2022temperature}  & Li$_3$ClO  
&2022 & Wang  \cite{wang2022thermal}  & Wadsleyite  \\
&2022 & Yang  \cite{yang2022lattice}  & MgSiO$_3$ perovskite 
&2022 & Zhang  \cite{zhang2022phonon} & Bi$_2$Te$_3$  \\
&2023 & Bhatt \cite{bhatt2023transition} & Tungsten 
&2023 & Dong \cite{dong2023development} & NaCl-MgCl$_2$-CaCl$_2$ \\
&2023 & Fu  \cite{fu2023medium}  & Ti-Zr-Y-Si-O ceramic 
&2023 & Gupta \cite{gupta2023distinct} & Bulk MoSe$_2$ and WSe$_2$ \\
&2023 & Han  \cite{han2023lattice}  & 2D InSe 
&2023 & Huang  \cite{huang2023grain}   & CdTe \\
&2023 & Li  \cite{li2023thermal,li2023thermal02}   & Cu/H$_2$O and TiO$_2$/H$_2$O 
&2023 & Qi  \cite{qi2023reversible}   & Vitreous silica \\
&2023 & Qiu \cite{qiu2023anomalous}  & Ice 
&2023 & Qu  \cite{qu2023deep}  & MnBi$_2$Te$_4$, Bi$_2$Te$_3$/MnBi$_2$Te$_4$ \\
&2023 & Ren  \cite{ren2023extreme}  & Ag$_8$SnSe$_6$ 
&2023 & Wang \cite{wang2023thermal} & Bridgmanite, Post-perovskite \\
&2023 & Xu  \cite{xu2023development}  & MgCl$_2$-NaCl and MgCl$_2$-KCl  
&2023 & Zhang  \cite{zhang2023tuning, zhang2023accessing}  & Sb$_2$Te$_3$ \\
&2023 & Zhang  \cite{Zhang2023thermal}  & Water 
&2023 & Zhang  \cite{zhang2023vacancy} & Boron arsenide \\
&2023 & Zhao  \cite{zhao2023microstructure}  & NaCl and NaCl-SiO$_2$  
&2024 & Fu    \cite{fu2024determining} & SiC \\
&2024 & Li    \cite{li2024deep} & AlN 
&2024 & Li \cite{Li2024acsami} & GaN/SiC interfaces\\
&2024 & Peng \cite{Peng2024grl} & MgSiO$_3$-H$_2$O
&2024 & Zhang    \cite{zhang2024dp} & MoAlB \\

%%%%%%%%%%%%%%%%%%%%%%%%%%%%%%%%% VASP-MLFF
\hline
MLFF & 2021 & Verdi   \cite{verdi2021npjvasp}   & ZrO$_2$  
     & 2024 & Lahnsteiner  \cite{Lahnsteiner2024jpcc}    & CsPbBr$_3$ \\

%%%%%%%%%%%%%%%%%%%%%%%%%%%%%%%%% NEP
\hline
\gls{nep} & 2021 & Fan  \cite{fan2021neuroevolution}   & PbTe, Si  
 &2022 & Dong  \cite{dong2022exactly}    & 2D silicene \\
 &2022 & Fan  \cite{fan2022jpcm,fan2022gpumd}    & PbTe, Amorphous carbon 
 &2023 & Cheng  \cite{cheng2023lattice}    & PbTe \\
 &2023 & Dong  \cite{dong2023ijhmt}    & C$_{60}$ network  
 &2023 & Du  \cite{du2023low}    & PH$_4$AlBr$_4$ \\
 &2023 & Eriksson  \cite{eriksson2023tuning}   & Graphite, $h$-BN, MoS$_2$ 
&2023 & Liang  \cite{ting2023prb}   & Amorphous SiO$_2$ \\
 &2023 & Liu  \cite{liu2023modulation} & Si/Ge nanowires 
 &2023 & Lu  \cite{lu2023reduction} & Fullerene-encapsulated CNT \\
 &2023 & Ouyang  \cite{ ouyang2023role} & AgX (X=Cl, Br, I) 
 &2023 & Pan  \cite{pan2023magnesium}   & MgOH system \\
 &2023 & Sha  \cite{sha2023phonon}     & 2D PbTe  
 &2023 & Shi  \cite{shi2023investigation1}    & InGeX$_3$ (X=S,Se,Te) \\
 &2023 & Shi  \cite{shi2023investigation2}    & Halogen perovskites 
 &2023 & Su  \cite{su2023origin}   & Cs$_2$BiAgBr$_6$, Cs$_2$BiAgCl$_6$ \\
 &2023 & Sun  \cite{sun2023neu}  & Ga$_2$O$_3$ 
 &2023 & Wang  \cite{wang2023quantum}   & Amorphous silicon \\
 &2023 & Wang  \cite{wang2023phonon}   & 2D SrTiO$_3$ 
 &2023 & Xu  \cite{xu2023accurate}   & Water \\
 &2023 & Xiong  \cite{xiong2023molecular}    & Diamond allotropes 
 &2023 & Ying  \cite{ying2023sub,ying2023ijhmt}   & MOF crystals, Phosphorene \\ 
 &2023 & Zhang  \cite{zhang2023prb}   & Amorphous HfO$_2$ 
 & 2024 & Cao \cite{Cao2024ijhmt} & Phosphorous carbide\\
 & 2024 & Cheng \cite{Cheng2024prb} & Perovskites
 &2024  & Fan \cite{fan2024anomalous} & HKUST-1 crystal \\
 &2024  & Fan \cite{fan2024combining}  & Graphene antidot lattice 
 &2024 & Li  \cite{Li-strained-graphene-PRB-2024} & Strained monolayer graphene\\
 &2024 & Li  \cite{li2024enhanced}   & Amorphous silicon 
 &2024 & Li  \cite{Li2024ijhmt}   & 2D COF-5\\
 & 2024 & Pegolo \cite{Pegolo2024FM} & Glassy Li$_x$Si$_{1-x}$ 
 &2024 & Wang  \cite{wang2023dissimilar}   & Ga$_2$O$_3$ \\
 &2024 & Ying  \cite{ying2023combining}   & MOF crystals 
 &2024 & Yue \cite{Yue2024prb} & Si-C interfaces \\
 &2024 & Zeraati  \cite{Zeraati2024prm}   & La$_2$Zr$_2$O$_7$ and many others 
 &2024 & Zhang  \cite{zhang2024thermal}   & GeTe \\
%%%%%%%%%%%%%%%%%%%%%%%%%%%%%%%%% So3krates
\hline
So3krates & 2023 & Langer  \cite{Langer2023jcp}   & SnSe\\ 
\hline
\hline
\end{tabular}
\end{center}
\label{table:mlp-kappa}
\end{table*}

The application of \glspl{mlp}-based \gls{md}  simulations to thermal transport was pioneered by Sosso \textit{et al.} in 2012 when they studied the thermal transport in the phase-changing amorphous GeTe system \cite{Sosso2012prb}. However, thermal transport simulations are very computationally intensive, and the rapid increase of the number of applications has only been started after the development of the GPU-based \gls{dp} \cite{Wang2018cpc} and \gls{nep} \cite{fan2021neuroevolution}  models. In this regard, the \gls{nep} model is particularly advantageous due to its superior computational speed as compared to others \cite{fan2021neuroevolution,fan2022jpcm,fan2022gpumd}. With comparable computational resources, it has been shown to be as fast as or even faster than some empirical force fields \cite{xu2023accurate,ying2023sub}.

\begin{figure}[htb]
\begin{center}
\includegraphics[width=\columnwidth]{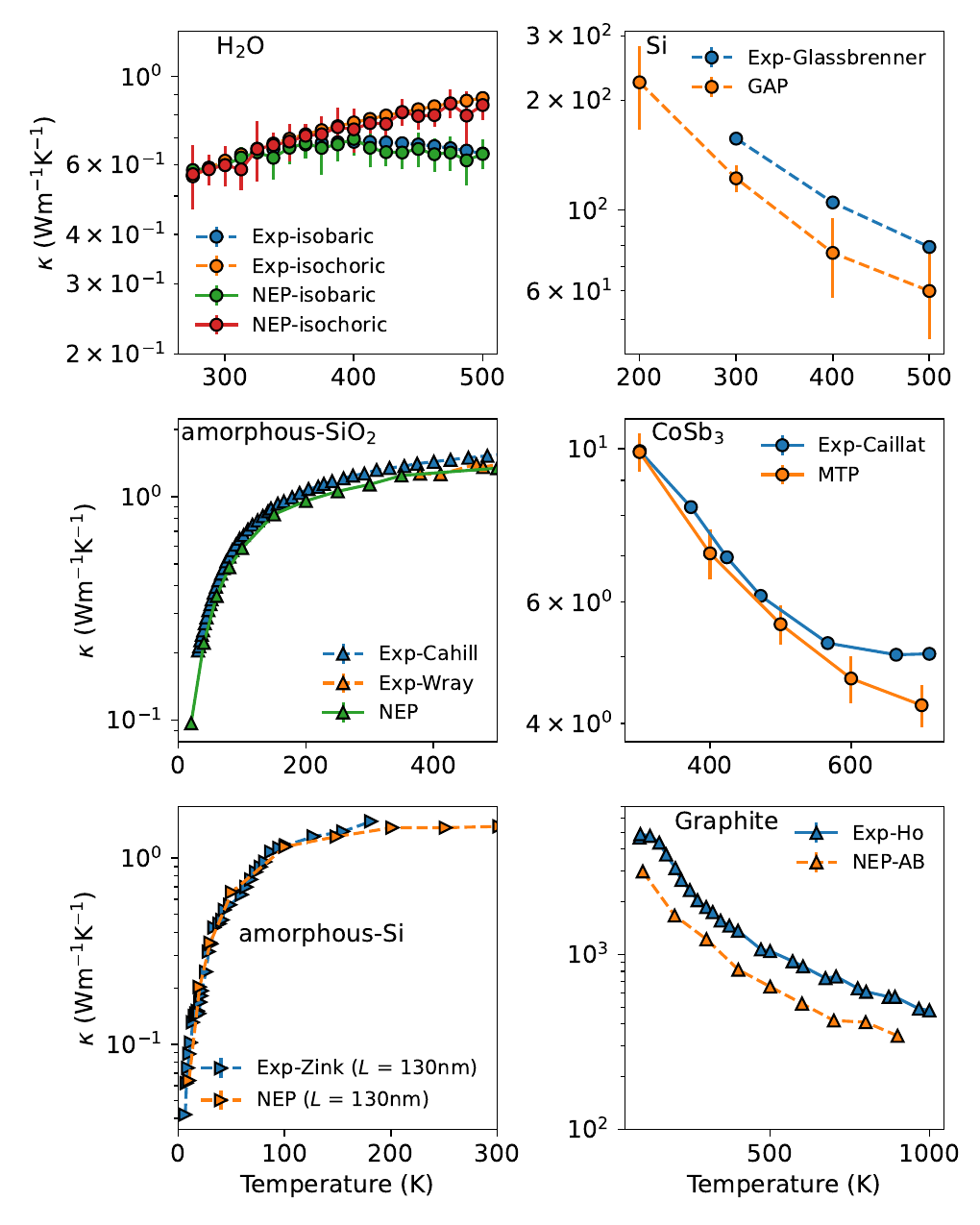}
\caption{Selected literature results on the application of \glspl{mlp} to thermal transport, covering a broad range of materials, including liquid water \cite{xu2023accurate}, amorphous SiO$_2$ \cite{ting2023prb}, amorphous silicon \cite{wang2023quantum}, crystalline silicon \cite{qian2019mtp},  crystalline CoSb$_3$ \cite{korotaev2019prb}, and crystalline graphite \cite{eriksson2023tuning}. Experimental data are from Refs. ~\onlinecite{huber2012new, Linstrom2022NIST} (liquid water), Refs.~\onlinecite{cahill1989heat,cahill1990thermal,wray1959thermal} (amorphous SiO$_2$), Ref.~\onlinecite{zink2006prl} (amorphous silicon), Ref.~\onlinecite{Glassbrenner1964Thermal} (crystalline silicon), Ref.~\onlinecite{caillat1996jap} (crystalline CoSb$_3$), and Ref.~\onlinecite{ho1972thermal} (crystalline graphite).}
\label{figure:review}
\end{center}
\end{figure}

There are numerous successful applications of \glspl{mlp} in thermal transport. In Fig.~\ref{figure:review}, we present results from selected publications. 
The materials studied in these works have reliable experimental results, serving as good candidates for validating the applicability of \glspl{mlp}. 
On one hand, \glspl{mlp} demonstrate good agreement with experimental results for highly disordered materials such as liquid water \cite{xu2023accurate}, amorphous SiO$_2$ \cite{ting2023prb}, and amorphous silicon \cite{wang2023quantum}. 
In addition to the reliability of \glspl{mlp}, a crucial component for accurately describing the temperature dependence of the thermal conductivity in liquids and amorphous materials is a quantum correction method based on the spectral thermal conductivity, as defined in Eq.~(\ref{equation:kappa_omega}), and the quantum-statistical-correction factor, as given in Eq.~(\ref{equation:quantum-correction-factor}).
On the other hand, \glspl{mlp} tend to systematically underestimate the thermal conductivity of crystalline solids, including silicon (using a \gls{gap} model)  \cite{qian2019mtp}, CoSb$_3$ (using a \gls{mtp} model), and graphite (in-plane transport, using a \gls{nep} model) \cite{eriksson2023tuning}.
This underestimation has been attributed to the small but finite random force errors, and a correction has been devised \cite{Wu2024correcting}. We will discuss this in more detail with an example in the next section.

\section{Molecular dynamics simulation of heat transport using NEP and GPUMD}

In this section, we use crystalline silicon as an example to demonstrate the workflow of constructing and using \gls{nep} models for thermal transport simulations. The \gls{nep} approach has been implemented in the open-source \textsc{gpumd} package \cite{fan2017cpc,fan2022gpumd}. After compiling, there will be an executable named \verb"nep" that can be used to train \textit{accurate} \gls{nep} models against reference data, and an executable named \verb"gpumd" that can be used to perform \textit{efficient} \gls{md} simulations. The \textsc{gpumd} package is self-contained, free from dependencies on third-party packages, particularly those related to \gls{ml}. This makes the installation of \textsc{gpumd} straightforward and effortless. In addition, there are some handy (but not mandatory) Python packages available to facilitate the pre-processing and post-processing \textsc{gpumd} inputs and outputs, including \textsc{calorine} \cite{Lindgren2024joss}, \textsc{gpyumd} \cite{gpyumd}, \textsc{gpumd-wizard} \cite{gpumd-wizard}, and \textsc{pynep} \cite{pynep}. Since its inception with the very first version in 2013\cite{fan2013cpc}, \textsc{gpumd} has been developed with special expertise in heat transport applications. 

\subsection{The neuroevolution potential}

The \gls{nep} model is based on \gls{ann} and is trained using a \gls{snes} \cite{Schaul2011High}, hence the name. 

\subsubsection{The NN model}

The \gls{ml} model in \gls{nep} is a fully-connected feedforward \gls{ann} with a single hidden layer, which is also called a multilayer perceptron. The total energy is the sum of the site energies $U=\sum_i U_i$, and the site energy $U_i$ is the output of the \gls{nn}, expressed as:
\begin{equation}
\label{equation:u_i}
U_{i} = \sum_{\mu=1}^{N_{\rm neu}} \omega_\mu^{(1)} \tanh\left(\sum_{\nu=1}^{N_{\rm des}}\omega_{\mu\nu}^{(0)}q_{\nu}^i-b_{\mu}^{(0)}\right)-b^{(1)}. 
\end{equation}
Here, $N_{\rm des}$ is the number of descriptor components, $N_{\rm neu}$ is the number of neurons in the hidden layer, $q_{\nu}^i$ is the $\nu$-th descriptor component of atom $i$, $\omega_{\mu\nu}^{(0)}$ is the connection weight matrix from the input layer to the hidden layer, $\omega_\mu^{(1)}$ is the connection weight vector from the hidden layer to the output layer, $b_{\mu}^{(0)}$ is the bias vector in the hidden layer, and $b^{(1)}$ is the bias in the output layer. $\omega_{\mu\nu}^{(0)}$, $\omega_\mu^{(1)}$, $b_{\mu}^{(0)}$, and $b^{(1)}$ are trainable parameters. The function $\tanh(x)$ is the nonlinear activation function in the hidden layer. According to Eq. (\ref{equation:u_i}), the \gls{nep} model is a simple analytical function of a descriptor vector. A C++ function for evaluating the energy and its derivative with respect to the descriptor components can be found in Ref.~\onlinecite{fan2022gpumd}.

\subsubsection{The descriptor}

The descriptor $q_i^{\nu}$ encompasses the local environment of atom $i$. In  \gls{nep}, the descriptor is an abstract vector whose components group into radial and angular parts.
The radial descriptor components $q_n^i $ $(0 \leq {n} \leq n_{\rm max}^{\rm R})$ are defined as
\begin{equation}
\label{equation:rad_des}
q_{n}^i = \sum_{j\neq{i}}g_n(r_{ij}), 
\end{equation}
where $r_{ij}$ is the distance between atoms $i$ and $j$ and $g_n(r_{ij})$ are a set of radial functions, each of which is formed by a linear combination of Chebyshev polynomials. The angular components include $n$-body ($n= 3,4,5$) correlations. For the 3-body part, the descriptor components are defined as $(0 \leq {n} \leq n_{\rm max}^{\rm A}$,  $1 \leq {l} \leq l_{\rm max}^{\rm 3body})$ 
\begin{equation}
\label{equation:ang_des}
q_{nl}^i = \sum_m (-1)^m A_{nlm}^i A_{nl(-m)}^i;
\end{equation}
\begin{equation}
A_{nlm}^i = \sum_{j\neq i} g_n(r_{ij}) Y_{lm}(\hat{\bm{r}}_{ij}).
\end{equation}
Here, $Y_{lm}$ are the spherical harmonics and $\hat{\bm{r}}_{ij}$ is the unit vector of $\bm{r}_{ij}$. Note that the radial functions $g_n(r_{ij})$ for the radial and angular descriptor components can have different cutoff radii, which are denoted as $r_{\rm c}^{\rm R}$ and $r_{\rm c}^{\rm A}$, respectively. For 4-body and 5-body descriptor components (with similar hyperparameters $l_{\rm max}^{\rm 4body}$ and $l_{\rm max}^{\rm 5body}$ as in the 3-body part), see Ref.~\onlinecite{fan2022gpumd}.

\subsubsection{The training algorithm}

The free parameters are optimized using the \gls{snes} by minimizing a loss function that is a weighted sum of the \glspl{rmse} of energy, force, and virial stress, over $N_{\rm gen}$ generations with a population size of $N_{\rm pop}$. The weights for the energy, force, and virial are denoted $\lambda_{\rm e}$, $\lambda_{\rm f}$, and $\lambda_{\rm v}$, respectively. Additionally, there are proper norm-1 ($\ell_1$) and norm-2 ($\ell_2$) regularization terms. For explicit details on the training algorithm, refer to Ref.~\onlinecite{fan2021neuroevolution}.

\subsubsection{Combining with other potentials}

Although \gls{nep} with proper hyperparameters can account for almost all types of interactions, it can be useful to combine it with some well developed potentials, such as the \gls{zbl} \cite{Ziegler1985} potential for describing the extremely large screened nuclear repulsion at short interatomic distances and the D3 dispersion correction \cite{Grimme2011jcc} for describing relatively long-range but weak interactions. Both potentials have been recently added to the \textsc{gpumd} package \cite{Liu2023prb,ying2023combining}
It has been demonstrated that dispersion interactions can reduce the thermal conductivity of typical metal-organic frameworks by about 10\% \cite{ying2023combining}. With the addition of \gls{zbl} and D3, \gls{nep} can then focus on describing the medium-range interactions. 

\subsection{Model training and testing}

There are educational articles focusing on various best practices in constructing \glspl{mlp} \cite{Miksch2021Strategies,Tokita2023jcp}. 
Here we use crystalline silicon as a specific example to illustrate the particular techniques in the context of \gls{nep}. 

\subsubsection{Prepare the initial training data}

A training dataset is a collection of structures, each characterized by a set of attributes:
\begin{enumerate}
    \item a cell matrix defining a periodic domain
    \item the species of the atoms in the cell
    \item the positions of the atoms
    \item the total energy of the cell of atoms
    \item the force vector acting on each of the atoms
    \item (optionally) the total virial tensor (with 6 independent components) of the cell
\end{enumerate}
The structures can be prepared by any method, while the energy, force, and virial are usually calculated via quantum mechanical methods, such as the \gls{dft} method.
For a dataset comprising $N_{\rm str}$ structures with a total number of $N$ atoms, there are $N_{\rm str}$ energy data, $6N_{\rm str}$ virial data, and $3N$ force data. 

While there are already several publicly available training datasets for silicon, we opt to create one from scratch for pedagogical purposes.
The construction of training dataset typically involves an iterative process, employing a  scheme similar to active learning. 
The iterative process begins with an initial dataset. To investigate heat transport in crystalline silicon, the initial training dataset should encompass structures relevant to the target temperatures and pressures.
The most reliable way of generating structures under these conditions is through performing  \gls{aimd} simulations, where interatomic forces are calculated based on quantum mechanical methods, such as the \gls{dft} approach.
However, \gls{aimd} is computationally expensive (which is the primary motivation for developing a \gls{mlp}) and it is often impractical to perform \gls{aimd} simulations for a dense grid of thermodynamic conditions. 
Fortunately, there is usually no such need for the purpose of generating the reference structures. 
Actually, manual perturbation of the atomic positions and/or the cell matrices proves to be an effective way of generating useful reference structures. 

Based on the considerations above, we generate the initial training dataset through the following methods. 
Firstly, we generate 50 structures by applying random strains (ranging from $-3\%$ to $+3\%$ for each degree of freedom) to the unit cell of cubic silicon (containing 8 atoms) while \textit{simultaneously} perturbing the atomic positions randomly  (by 0.1 \AA). 
Secondly, we perform a 10-ps \gls{aimd} simulation at 1000 K (fixed cell) using a $2 \times 2 \times 2$ supercell of silicon crystal containing 64 atoms, and sample the structures every 0.1 ps, obtaining another 100 structures. In total, we obtain 150 structures and 6800 atoms initially.

After obtaining the structures, we perform single-point \gls{dft} calculations to obtain the reference energy, force and virial data. These data are saved to a file named \verb"train.xyz", using the extended XYZ format. The single-point \gls{dft} calculations are performed using the 
\textsc{vasp} package \cite{Blochl1994Projector}, using the Perdew-Burke-Ernzerhof functional with the generalized gradient approximation \cite{Perdew1997Generalized}, a cutoff energy of 600 eV, an energy convergence threshold of $10^{-6}$ eV, and a $k$-point mesh of $4\times4\times4$ for 64-atom supercells and $12\times12\times12$ for 8-atom unit cells.

\subsubsection{Train the first NEP model}

With the training data, we proceed to train our first \gls{nep} model, denoted as \gls{nep}-iteration-1. For this task, we need to prepare an input file named \verb"nep.in" for the \verb"nep" executable in the \textsc{gpumd} package. This \verb"nep.in" input file contains the various hyperparameters for the \gls{nep} model under training. Most hyperparameters have well-suited default values, and for users  initiating this process, it is recommended to use these defaults whenever applicable. The default values for key hyperparameters are as follows:
\begin{enumerate}
\item $n_{\rm max}^{\rm R}$: 4
\item $n_{\rm max}^{\rm A}$: 4
\item Chebyshev polynomial basis size for radial descriptor components $N^\mathrm{R}_\mathrm{bas}$: 12
\item Chebyshev polynomial basis size for angular descriptor components $N^\mathrm{A}_\mathrm{bas}$: 12
\item $l_{\rm max}^{\rm 3body}$: 4
\item $l_{\rm max}^{\rm 4body}$: 2
\item $l_{\rm max}^{\rm 5body}$: 0 (not used by default)
\item $N_{\rm neu}$: 30
\item Energy and force weights $\lambda_{\rm e}$ and $\lambda_{\rm f}$: 1 
\item Virial weight $\lambda_{\rm v}$: 0.1 
\item Batch size: 1000 (a large or even full batch is preferred for training with \gls{snes})
\item Population size in \gls{snes}: 50
\item Number of training generations (steps): \num{100000}
\item \gls{ann} architecture: 30-30-1 (input layer 30, hidden layer 30, scalar output; relatively small but sufficient for most cases, expect for very complicated training data.)
\end{enumerate} 

Following this strategy, we use a very simple \verb"nep.in" input file for our case, which is as follows:
\begin{Verbatim}[frame=single]
type        1  Si
cutoff      5  5
\end{Verbatim}
In the first line, we specify the number of species (atom types) and the chemical symbol(s). In our example, there is only one species with the chemical symbol Si. In the second line, we specify the cutoff radii $r_{\rm c}^{\rm R}$ and $r_{\rm c}^{\rm A}$ for the $g_n(r_{ij})$ functions in the radial and angular descriptor components, respectively. In our example, both cutoff radii are set to 5 \AA{}, which includes the third nearest neighbors. The choice of cutoff radii is crucial for the performance of the trained \gls{nep} model and usually requires a systematic exploration to find an optimal set of values. It is important to note that the average number of neighbors, and hence the computational cost, scales cubically with respect to the cutoff radii. Therefore, blindly using large cutoff radii is not advisable. Although $r_{\rm c}^{\rm R}=r_{\rm c}^{\rm A}$ in our current example, it is generally beneficial to use a larger $r_{\rm c}^{\rm R}$ and a smaller $r_{\rm c}^{\rm A}$,  because the radial descriptor components are computationally much cheaper than the angular descriptor components. Using a larger $r_{\rm c}^{\rm R}$ does not lead to a significant increase in the computational cost, but can help capture longer-range interactions (such as screened Coulomb interactions in ionic compounds \cite{fan2021neuroevolution}) that typically have little angular dependence. A larger radial cutoff is also useful for capturing dispersion interactions in Van der Waals structures \cite{eriksson2023tuning}.

\begin{figure}[htb]
\begin{center}
\includegraphics[width=\columnwidth]{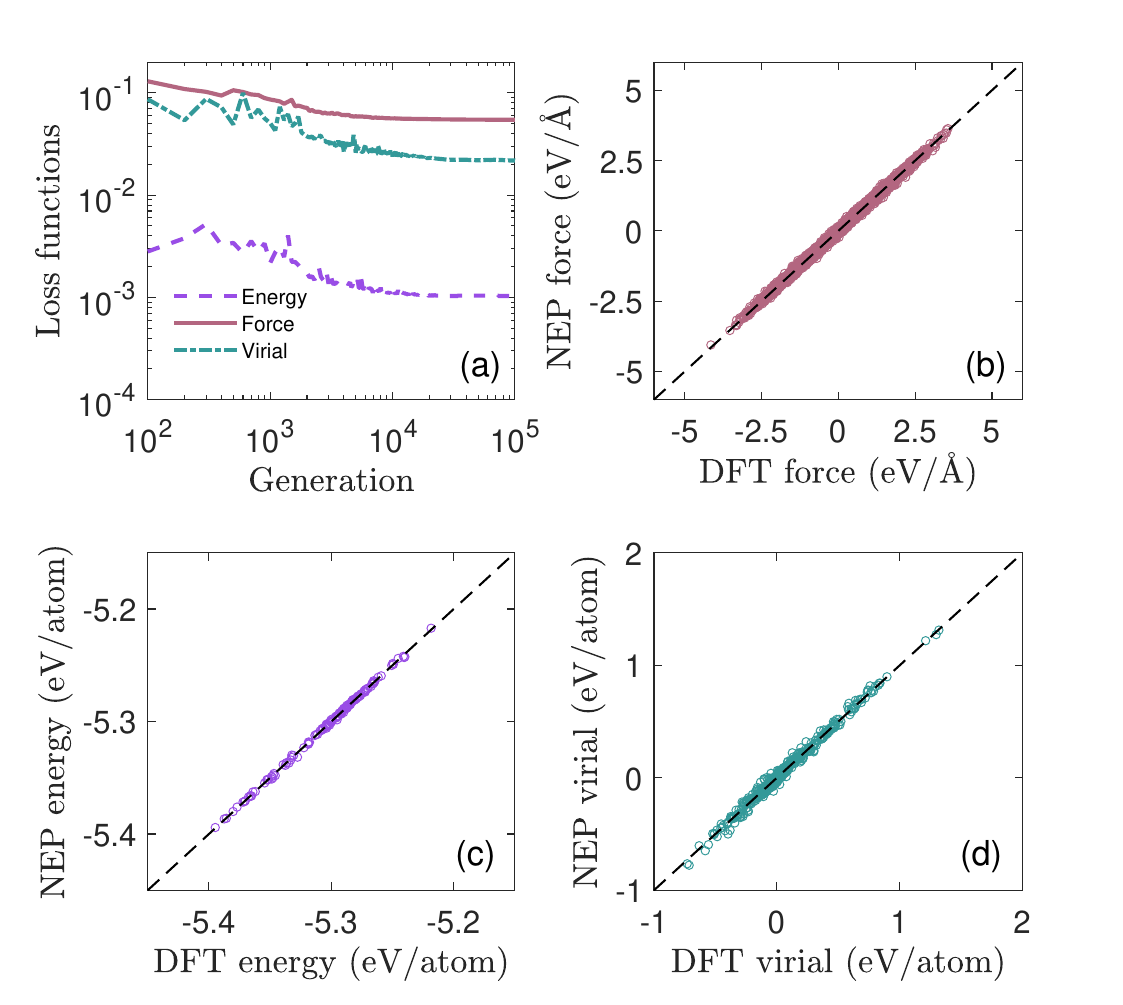}
\caption{(a) Evolution of \glspl{rmse} of energy, force, and virial with respect to training generations (steps). (b) Comparison of force, (c) energy, and (d) virial calculated by \gls{nep} against \gls{dft} reference data for the initial training dataset.}
\label{figure:train-v1}
\end{center}
\end{figure}

The training results for NEP-iteration-1 are shown in Fig. \ref{figure:train-v1}(a). The \glspl{rmse} of force, energy, and virial all converge well within the default \num{100000} training steps. The parity plots for force, energy, and virial in Figs. \ref{figure:train-v1}(b)-\ref{figure:train-v1}(d) show good correlations between the \gls{nep} predictions and the \gls{dft} reference data. The \glspl{rmse} for energy, force, and virial are 1.0 meV/atom, 54.6 meV/\AA{}, and 21.8 meV/atom, respectively . 

\subsubsection{Training iterations}

Reliable assessment of the accuracy of a \gls{mlp} typically involves an independent test dataset rather than the training dataset. To this end, we perform 10-ps \gls{md} simulations using \gls{nep}-iteration-1 in the $NPT$ ensemble. The target pressure is set to zero, and the target temperatures range from 100 K to 1000 K with  intervals of 100 K. We sample 100 structures, totalling 6400 atoms.

We perform single-point \gls{dft} calculations for these structures and then use \gls{nep}-iteration-1 to generate predictions. This is achieved by adding the \verb"prediciton" keyword to the  \verb"nep.in" file:
\begin{Verbatim}[frame=single]
type        1  Si
cutoff      5  5
prediction  1
\end{Verbatim}
This results in a rapid prediction for the test dataset. The \glspl{rmse} for energy, force, and virial are 1.2 meV/atom, 41.6 meV/\AA{}, and 8.5 meV/atom, respectively. These values are already comparable to those for the training dataset, indicating that we can actually stop here and use \gls{nep}-iteration-1 as the final model.
However, for added confidence, it is generally advisable to perform at least one more iteration. Therefore, we combine the test dataset (100 structures) with the training dataset (150 structures) to form an expanded training dataset (250 structures), and then train a new model named \gls{nep}-iteration-2.
With this new \gls{nep} model, we generate another test dataset with 100 structures, using similar procedure as above but with a simulation time of 10 ns (instead of 10 ps), driven by \gls{nep}-iteration-2 for each temperature. The test \glspl{rmse} for \gls{nep}-iteration-2 are 0.5 meV/atom (energy), 33.5 meV/\AA{} (force), and 8.9 meV/atom (virial), respectively. Both the energy and force \glspl{rmse} are smaller than those for the previous iteration, indicating the improved performance of \gls{nep}-iteration-2 compared to \gls{nep}-iteration-1.

The high accuracy of the latest test dataset sampled from 10-ns \gls{md} simulations driven by \gls{nep}-iteration-2 suggests that \gls{nep}-iteration-2 is a reliable model for \gls{md} simulation of crystalline silicon from 100 to 1000 K. Therefore, we conclude the iteration and use \gls{nep}-iteration-2 for the thermal transport applications. In the following, we will refer to \gls{nep}-iteration-2 simply as \gls{nep}.
This \gls{nep} model, running on a consumer-grade NVIDIA RTX 4090 \gls{gpu} card with 24 GB of memory, achieves a remarkable computational speed of about $2.4\times 10^7$ atom-step/second, equivalent to a computational cost of about $4.2\times 10^{-8}$ s/atom/step in \gls{md} simulations.

Using a trained \gls{mlp} to generate \gls{md} trajectory is a common practice in nearly all the active-learning schemes documented in the literature. The major difference between different active-learning schemes is about the criteria for selecting structures to be added to the training dataset. While there might be a risk of sampling nonphysical structures using a trained \gls{mlp} model, as demonstrated in this tutorial, one can mitigate the risk by conducting a few iterations and employing shorter \gls{md} runs in the initial stages, progressively increasing the \gls{md} simulation time with each iteration. As a result, the \gls{mlp} becomes increasingly reliable throughout the iteration process, enabling the generation of longer and more accurate trajectories over time. In our example using the silicon crystal, a relatively simple system, we only performed two iterations to achieve accurate predictions for 10-ns \gls{md} runs. However, for more complex systems, one might need to perform more iterations, increasing the \gls{md} steps more gradually than what we have demonstrated for the silicon crystal example.

\subsection{Thermal transport applications}

\subsubsection{Phonon dispersion relations}

\begin{figure}[htb]
\begin{center}
\includegraphics[width=\columnwidth]{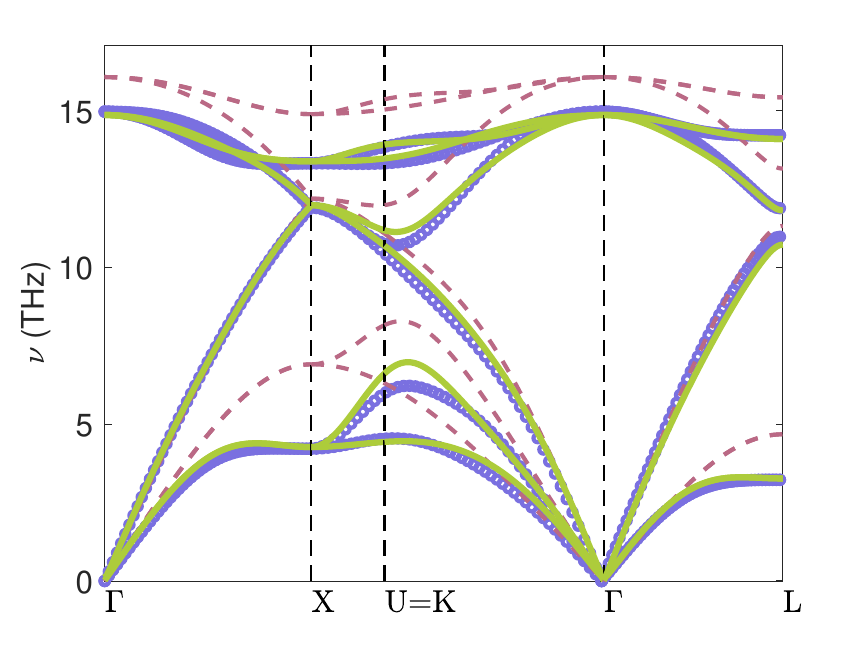}
\caption{ Phonon dispersion relations of silicon from \gls{dft} (circles), Tersoff potential (dashed lines), and NEP (solid lines).}
\label{figure:phonon}
\end{center}
\end{figure}

Before applying a \gls{mlp} to thermal transport applications, it is usually a good practice to examine the phonon dispersion relations. 
The phonon dispersion relations for \gls{nep} and Tersoff \cite{tersoff1989prb} potentials are calculated using \textsc{gpumd}, employing the finite-displacement method with a displacement of 0.01 \AA.
For \gls{dft}, we use density functional perturbation theory as implemented in \textsc{vasp} in combination with \textsc{phonopy} \cite{togo2015sm}, using a $4\times4\times4$ supercell, a cutoff energy of 600 eV, an energy convergence threshold of $10^{-8}$ eV, and a $5\times5\times5$ $k$-point mesh.

In Fig.~\ref{figure:phonon}, we compare the phonon dispersion relations  calculated from \gls{dft}, Tersoff potential, and \gls{nep}.
While there are small differences between \gls{nep} and \gls{dft} results, the agreement between \gls{nep} and \gls{dft} is significantly better than that between Tersoff and \gls{dft}. The agreement between \gls{nep} and \gls{dft} can, in principle, be further improved, for example, by increasing the size of the \gls{ann} model and/or the cutoff radii. However, this comes with a trade-off, as it may reduce computational efficiency. In practice, achieving a  balance between accuracy and speed is essential.

\subsubsection{Thermal conductivity from EMD}

After validating the phonon dispersion relations, we proceed to thermal conductivity calculations using the various \gls{md} methods, as reviewed in Sec.~\ref{section:MD-methods}. All calculations are performed using the \verb"gpumd" executable in \textsc{gpumd}.

We start with the \gls{emd} method, using a sufficiently large $12\times12\times12$ cubic supercell with \num{13824} atoms. The \verb"run.in" file for the \verb"gpumd" executable is configured as follows:
\begin{Verbatim}[frame=single, fontsize=\small]
potential     nep.txt
velocity      300

ensemble      npt_ber 300 300 100 0 53.4059 2000
time_step     1     
dump_thermo   1000     
run           500000

ensemble      nve
compute_hac   20 50000 10
run           10000000
\end{Verbatim}
There are three input blocks. In the first block, we specify the \gls{nep} potential file and set the initial temperature to 300 K. The second block represents an equilibration run of 500 ps in the $NPT$ ensemble, aiming to reach a target temperature of 300 K and a target pressure of zero. The third block corresponds to a production run of 10 ns in the $NVE$ ensemble, with heat current sampled every 20 steps.

\begin{figure}[htb]
\begin{center}
\includegraphics[width=\columnwidth]{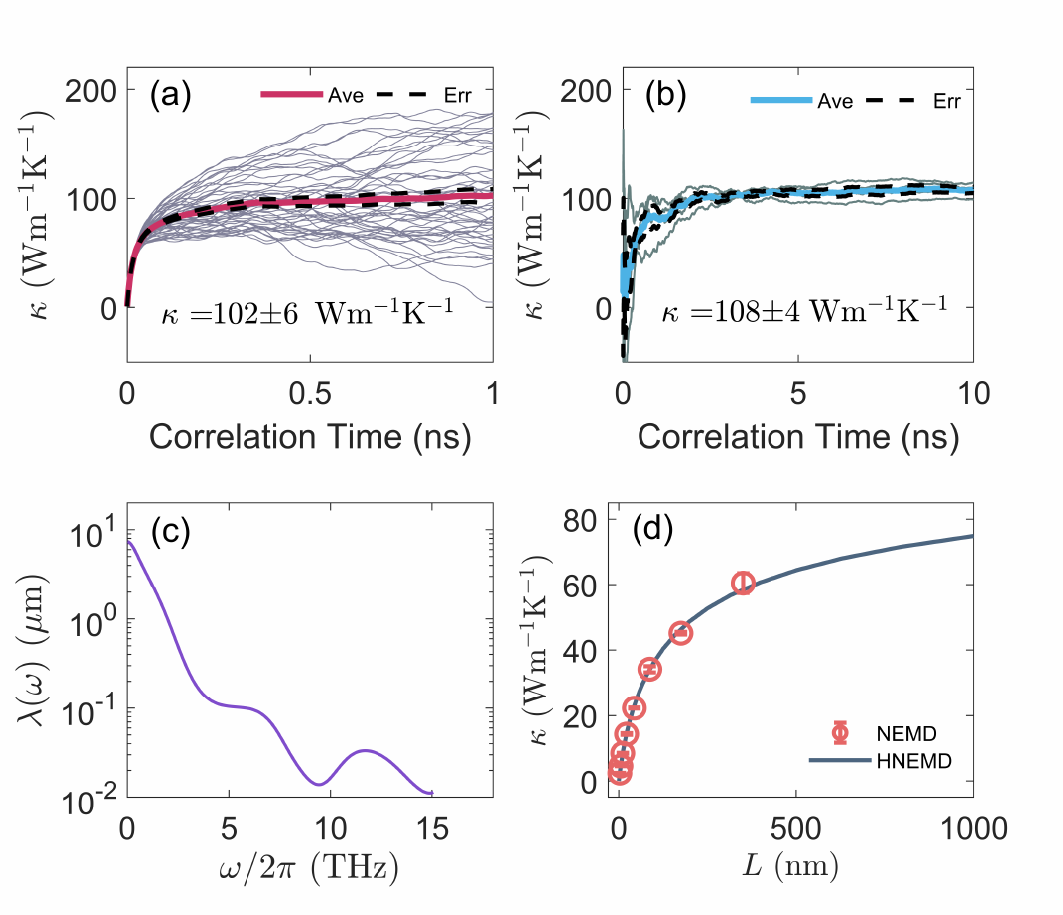}
\caption{ Thermal conductivity of crystalline silicon at 300 K  from three \gls{md}-based methods using the herein developed \gls{nep}. (a) Results from 50 independent \gls{emd} runs (thin solid lines), along with their average (thick solid line) and error bounds (thick dashed lines); (b) Results from 4 independent \gls{hnemd} runs (thin solid lines), along with their average (thick solid line) and error bounds (thick dashed lines); (c) Phonon \gls{mfp} spectrum calculated using spectral decomposition method; (d) Results from \gls{nemd} simulations (red symbols), matching the $\kappa(L)$ curve from the \gls{hnemd}-based formalism.}
\label{figure:kappa}
\end{center}
\end{figure}

We perform 50 independent runs using the inputs above, each with a different set of initial velocities. The $\kappa(t)$ [cf. Eq.~(\ref{equation:kappa_t_emd})] results from individual runs (thin solid lines) and their average (thick solid line) and error bounds (thick dashed lines) are shown in Fig.~\ref{figure:kappa}(a). Taking $t=1$ ns as the upper limit of the correlation time, up to which  $\kappa(t)$ converges well, we have $\kappa \approx 102 \pm 6$ Wm$^{-1}$K$^{-1}$ from the \gls{emd} method. In this work, all statistical errors are calculated as the standard error of the mean.

\subsubsection{Thermal conductivity from HNEMD}

We then move to the \gls{hnemd} method. Since the \gls{hnemd} method has the same finite-size effects as in the \gls{emd} method, we use the same simulation cell as in the \gls{emd} method. The \verb"run.in" file for the \verb"gpumd" executable reads as follows:
\begin{Verbatim}[frame=single, fontsize=\small]
potential     nep.txt
velocity      300

ensemble      npt_ber 300 300 100 0 53.4059 2000
time_step     1
dump_thermo   1000   
run           1000000

ensemble      nvt_nhc 300 300 100
compute_hnemd 1000  2e-5 0 0
compute_shc   2 250 0 1000 120
dump_thermo   1000
run           10000000
\end{Verbatim}
There are also three input blocks, and only the production block differs from the case of \gls{emd}. Here, the temperature is controlled using the Nose-Hoover chain thermostat, and an external force in the $x$ direction with $F_{\rm e}=2\times 10^{-5}$ \AA$^{-1}$ is applied. The production run has 10 ns in total. 

We perform 4 independent runs using the specified inputs, each with a different set of initial velocities. The $\kappa(t)$ [cf. Eq.~(\ref{equation:kappa_t_hnemd})] results from individual runs (thin solid lines) and their average (thick solid line) and error bounds (thick dashed lines) are shown in Fig.~\ref{figure:kappa}(b). The estimated thermal conductivity is  $\kappa \approx 108 \pm 4$ Wm$^{-1}$K$^{-1}$, consistent with the \gls{emd} value within statistical error bounds. It is noteworthy that the total production time for the \gls{hnemd} simulations ($4\times 10$ ns) is considerably smaller than that for the \gls{emd} simulations ($50\times 10$ ns), while the former still gives a smaller statistical error. This suggests a higher computational efficiency of the \gls{hnemd} over the \gls{emd} method, as previously emphasized \cite{fan2019prb}.

From the \gls{hnemd} simulations, we also obtain the spectral thermal conductivity $\kappa(\omega)$ [cf. Eq.~(\ref{equation:kappa_omega})]. By combining this with the spectral conductance $G(\omega)$ [cf. Eq.~(\ref{equation:G_omega})] in a ballistic \gls{nemd} simulation (details provided below), we calculate the phonon \gls{mfp} spectrum as 
\begin{equation}
\label{equation:lambda_omega}
    \lambda(\omega)= \frac{\kappa(\omega)}{G(\omega)},
\end{equation}
which is a generalization of Eq.~(\ref{equation:mfp_from_G}). The calculated $\lambda(\omega)$ is shown in Fig.~\ref{figure:kappa}(c). Remarkably, in the low-frequency limit, $\lambda(\omega)$ can go well beyond one micron.
With $\kappa(\omega)$ and $\lambda(\omega)$, one can calculate the spectral apparent thermal conductivity $\kappa(\omega,L)$ according to Eq.~(\ref{equation:ballistic_to_diffusive_lambda_omega}) and obtain the apparent thermal conductivity at \textit{any} length $L$ using Eq.~(\ref{equation:kappa_L_from_integration}).
The results are depicted by the solid line in Fig.~\ref{figure:kappa}(d).

\subsubsection{Thermal conductivity from NEMD\label{section:nemd-si}}

The third \gls{md} method we demonstrate is the \gls{nemd} method, using the fixed boundary setup discussed in Sec.~\ref{section:nemd}. We explore lengths $L=$ 2.7, 5.5, 11.0, 21.9, 43.8, 87.6, 175.3, 350.5 nm, maintaining a consistent $5 \times 5$ cell in the transverse direction. The heat source and sink regions span 4.4 nm, which is long enough to ensure fully thermalized phonons within these regions. The \verb"run.in" input file for our \gls{nemd} simulations reads as follows:
\begin{Verbatim}[frame=single, fontsize=\small]
potential     nep.txt
velocity      300

ensemble      nvt_ber 300 300 100
time_step     1 
fix           0
dump_thermo   100        
run           100000

ensemble      heat_lan 300 100 10 1 7 
fix           0
compute       0 10 100 temperature
compute_shc   2 250 0 1000 120.0 group 0 4
run           2000000
\end{Verbatim}

Unlike the \gls{emd} and \gls{hnemd} simulations, the \gls{nemd} simulations involve an extra operation: certain atoms are frozen.  We assign these atoms to the ``group'' 0 and use the \verb"fix 0" command to freeze them.
In the production stage, two Langevin thermostats with different temperatures are applied separately to groups 1 and 7, corresponding to the heat source and the heat sink, respectively. The temperature difference between them is set to 20 K. The heat flux can be obtained from the data produced by the \verb"compute" keyword, allowing us to calculate the apparent thermal conductivity $\kappa(L)$ according to Eq.~(\ref{equation:kappa_L_NEMD}). The production stage has a duration of 2 ns, with a well-established steady state achieved within the first 1 ns. Therefore, we use the second half of the production time to calculated the aforementioned stead-state properties. For each system length, we perform 2 independent runs, each with a different set of initial velocities. To get the  spectral conductance $G(\omega)$ in the ballistic limit, as used in Eq.~(\ref{equation:lambda_omega}), we use the data produced by the \verb"compute_shc" keyword in \gls{nemd} simulations with a short system length of $L=$ 1.6 nm.

\begin{figure}[htb]
% \begin{center}
\hspace{0.42cm}
\includegraphics[width=1.05\columnwidth]{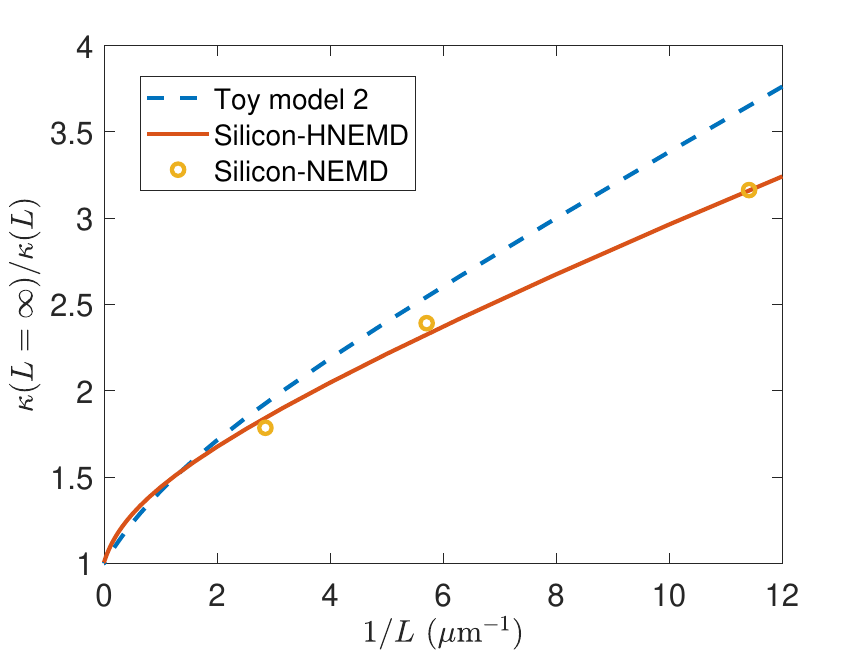}
\caption{The nonlinearity in the relation between $\kappa(L=\infty)/\kappa(L)$ and $1/L$ in the large-$L$ limit, observed in the second toy model (as discussed in Fig.~\ref{figure:ballistic_to_diffusive}(d)) and the silicon example.}
\label{figure:inverse-kappa-silicon}
% \end{center}
\end{figure}

As expected, the $\kappa(L)$ values from \gls{nemd} simulations match well with the $\kappa(L)$ curve from the \gls{hnemd}-based formalism [Fig.~\ref{figure:kappa}(d)]. 
However, reaching the diffusive limit directly through \gls{nemd} simulations is computationally demanding. Considering the presence of different phonon \glspl{mfp} [Fig.~\ref{figure:kappa}(c)] in the system, linear extrapolation to the diffusive limit based on a limited number of $\kappa(L)$ values from \gls{nemd} simulations is often inadequate. This limitation arises because the relation between $1/\kappa(L)$ and $1/L$ becomes nonlinear in the large-$L$ limit (see Fig.~\ref{figure:inverse-kappa-silicon}). This nonlinearity is a general feature in realistic materials, as also demonstrated in our toy model [Fig.~\ref{figure:ballistic_to_diffusive}(d)].

As of now, we have demonstrated the full consistency among the three \gls{md}-based methods. Notably, the \gls{hnemd} method stands out as the most computationally efficient. This explains why most works based on \textsc{gpumd} utilize the \gls{hnemd} method, with the other two methods typically being employed primarily for sanity-checking the results. 

\subsubsection{Comparison with experiments}

After obtaining consistent results from three \gls{md} methods, we are ready to compare the results with experimental data. The thermal conductivity of crystalline silicon is measured to be about 150 W m$^{-1}$ K$^{-1}$, but our \gls{hnemd} simulations predict a value of $108\pm 4$ W m$^{-1}$ K$^{-1}$, which is only $72\%$ of the experimental value. As a comparison, the thermal conductivity of crystalline silicon has been calculated \cite{dong2018prb} to be about $250 \pm 10$ W m$^{-1}$ K$^{-1}$ using a Tersoff potential \cite{tersoff1989prb}, which is $167\%$ of the experimental value. Specifically, the Tesoff potential appears to underestimate the phonon anharmonicity, while the \gls{nep} model tends to overestimate it. 

\begin{figure}[htb]
\begin{center}
\includegraphics[width=\columnwidth]{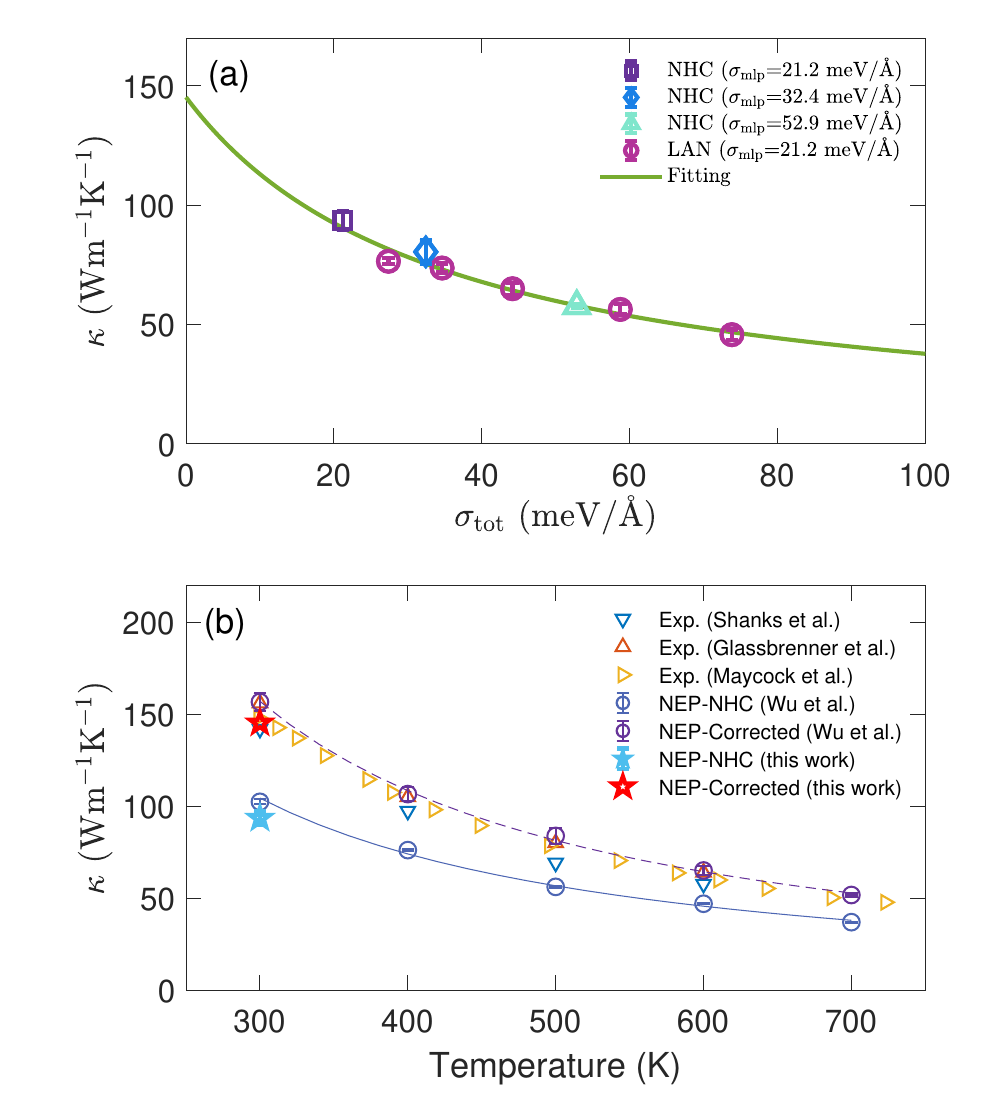}
\caption{ (a) Thermal conductivity of crystalline silicon at 300 K from \gls{hnemd} simulations using the herein developed \gls{nep} models as a function of the total force error $\sigma_{\rm tot}$. NHC and LAN represent the Nos\'{e}-Hoover chain and Langevin thermostatting methods, respectively. The data is fitted to obtain the corrected thermal conductivity of $\kappa(\sigma_{\rm tot}=0)=$ 145 W m$^{-1}$ K$^{-1}$. (b) Comparison of simulation results before and after the correction with experimental values \cite{Shanks1963Thermal, Glassbrenner1964Thermal, Maycock1967Thermal} and previously (uncorrected and corrected) \gls{nep}-\gls{md} simulations \cite{Wu2024correcting}.}
\label{figure:corrected}
\end{center}
\end{figure}

According to a recent unpublished study by Wu \textit{et al.}\cite{Wu2024correcting}, the underestimation of thermal conductivity by \glspl{mlp} could potentially be attributed to small but finite force errors compared to the reference data, leading to extra phonon scatterings. Based on the fact that the force errors form a Gaussian distribution, similar to the random forces in the Langevin thermostat, a method for correcting the force-error-induced underestimation of the thermal conductivity from \glspl{mlp} is proposed \cite{Wu2024correcting}. This correction involves conducting a series of \gls{hnemd} simulations with the temperature being controlled by a Langevin thermostat with various relaxation times $\tau_T$. Each component of the random force follows a Gaussian distribution with zero mean and a variance of 
\begin{equation}
\label{equation:sigma_L}
    \sigma^2_{\rm lan}= \frac{2k_{\rm B}Tm}{\tau_T \Delta t}, 
\end{equation}
where $m$ is the average atom mass in the system and $\Delta t$ is the integration time step. When the random forces in the Langevin thermostat and the force errors in the \gls{mlp} (with a \gls{rmse} of $\sigma_{\rm mlp}$ at a particular temperature) are present simultaneously, a new set of force errors is created, with a larger variance given by
\begin{equation}\label{equation:sum_sigma}
    {\sigma_{\rm tot}}^2={\sigma_{\rm lan}}^2+{\sigma_{\rm mlp}}^2,
\end{equation}
according to the properties of Gaussian distribution. After obtaining $\kappa(\sigma_{\rm tot})$ at different $\sigma_{\rm tot}$, the thermal conductivity with zero total force error $\kappa(\sigma_{\rm tot} = 0)$ can be obtained from the following relation \cite{Wu2024correcting}:
\begin{equation}
\label{equation:kappa_no_error}
\frac{1}{\kappa(\sigma_{\rm tot})} =\frac{1}{\kappa(\sigma_{\rm tot} = 0)} + \beta \sigma_{\rm tot},
\end{equation}
where $\beta$ is a fitting parameter.

Based on the correction method, we perform \gls{hnemd} simulations using the Langevin thermostat with the following set of $\tau_T$ values: 30, 50, 100, 200, and 500 ps. From these, the $\sigma_{\rm L}$ values are calculated to be 17.3, 27.4 38.7 54.8, and 70.7 meV/\AA. At 300 K, the force \gls{rmse} for our \gls{nep} model is tested to be $\sigma_{\rm mlp}=$ 21.2 meV/\AA. Therefore, the resulting $\sigma_{\rm tot}$ values are 27.4, 34.6, 44.2, 58.7, and 73.8 meV/\AA. To ensure consistency with experimental conditions, we also account for the presence of a few Si isotopes (92.2\% $^{28}$Si, 4.7\% $^{29}$Si, and 3.1\% $^{30}$Si) in the calculations. The calculated  $\kappa(\sigma_{\rm tot})$ with the various $\sigma_{\rm tot}$ are shown in Fig.~\ref{figure:corrected}(a). By fitting these data, we obtain a corrected thermal conductivity of $\kappa(\sigma_{\rm tot}=0)=$ 145 W m$^{-1}$ K$^{-1}$, in excellent agreement with the experimental value.

The extrapolation scheme described by Eq.~(\ref{equation:kappa_no_error}) not only applies to a single \gls{nep} model with different levels of intentionally added random forces through the Langevin thermostat, but is also valid for different \gls{nep} models with varying accuracy. To demonstrate this, we construct two extra \gls{nep} models with reduced accuracy. Starting from the default hyperparameters, we construct the first extra \gls{nep} model by reducing the number of neurons in the hidden layer from 30 to 1, resulting in an increased force \gls{rmse} of 32.4 meV/\AA. Based on this, we then construct the second extra \gls{nep} model by further reducing the Chebyshev polynomial basis sizes ($N^\mathrm{R}_\mathrm{bas}$, $N^\mathrm{A}_\mathrm{bas}$) from $(12, 12)$ to $(4, 4)$, resulting in a further increased force \gls{rmse} of 52.9 meV/\AA. The thermal conductivity results from the three \gls{nep} models with different accuracy using the Nos\'{e}-Hoover chain thermostat also closely follow the extrapolation curve [Fig.~\ref{figure:corrected}(a)], providing further support for the validity of the extrapolation scheme Eq.~(\ref{equation:kappa_no_error}).

Our results for 300 K before and after the correction are consistent with those reported in the previous work \cite{Wu2024correcting}, which also uses a \gls{nep} model [Fig.~\ref{figure:corrected}(b)]. In Fig.~\ref{figure:corrected}(b), we also show the results for other temperatures \cite{Wu2024correcting} in comparison to the experimental data. The corrected results agree well with the experimental ones across a broad range of temperatures. The slightly higher values from corrected \gls{nep} model predictions are likely due to the fact that isotope disorder was not considered in the previous calculations \cite{Wu2024correcting}.

While we have only demonstrated the application of the extrapolation (correction) method to \gls{hnemd} simulations, it is worth noting that this method is also potentially applicable to \gls{emd} simulations. We speculate that the force errors in \glspl{mlp} may also play a role in \gls{ald}-based approaches for thermal transport. 

\section{Summary and Conclusions \label{section:summary}}

In summary, we have provided a comprehensive pedagogical introduction to \gls{md} simulations of thermal transport utilizing the \gls{nep} \gls{mlp} as implemented in the \textsc{gpumd} package. 

We began by reviewing fundamental concepts related to thermal transport in both ballistic and diffusive regimes, elucidating the explicit expression of the heat flux in the context of \glspl{mlp}, and exploring various \gls{md}-based methods for thermal transport studies, including \gls{emd}, \gls{nemd}, \gls{hnemd}, and spectral decomposition. 

Following this, we conducted an up-to-date review of the literature on the application of \glspl{mlp} in thermal transport problems through \gls{md} simulations. 

A detailed review of the \gls{nep} approach followed, with a step-by-step demonstration of the process of developing an accurate and efficient \gls{nep} model for crystalline silicon applicable across a range of temperatures. Utilizing the developed \gls{nep} model, we explained the technical details of all \gls{md}-based methods for thermal transport discussed in this work. Finally, we compared the simulation results with experimental data, addressing the common trend of thermal conductivity underestimation by \glspl{mlp} and demonstrating an effective correction method.

By completing this tutorial, readers will be equipped to construct \glspl{mlp} and seamlessly integrate them into highly efficient and predictive \gls{md} simulations of heat transport.

\vspace{0.5cm}
\noindent \textbf{Data availability:}
All the training and test datasets and the trained \gls{nep} models for crystalline silicon are freely available at \url{https://gitlab.com/brucefan1983/nep-data}. 
The training datasets, trained \gls{nep}, \gls{dp}, and \gls{mtp} models for graphene and \gls{md} input files for reproducing Fig.~\ref{figure:heatflux} are freely available at \url{https://github.com/hityingph/supporting-info/tree/main/Dong_GPUMD_Tutorial_2024}. 

\begin{acknowledgments}
HD is supported by the Science Foundation from Education Department of Liaoning Province (No. JYTMS20231613) and the Doctoral start-up Fund of Bohai University (No. 0523bs008). PY is supported by the Israel Academy of Sciences and Humanities \& Council for Higher Education Excellence Fellowship Program for International Postdoctoral Researchers. KX and TL acknowledge support from the National Key R\&D Project from Ministry of Science and Technology of China (No. 2022YFA1203100), the Research Grants Council of Hong Kong (No. AoE/P-701/20), and RGC GRF (No. 14220022). ZZ acknowledges the European Union’s Horizon 2020 research and innovation programme under the Marie Skłodowska-Curie grant agreement No 101034413. SX acknowledges the financial support from the National Natural Science Foundation of China (Grant No.12174276). 

\end{acknowledgments}

\appendix

\section{Details on heat current validation \label{section:appendix}}

To validate the implementation of heat current for \gls{nep}, \gls{dp}, and \gls{mtp} [see Fig.~\ref{figure:heatflux}], we use a common reference dataset to train a model for each of the three \glspl{mlp}. We take all the monolayer graphene structures from Ref. \onlinecite{wen2019prb} and use \num{3288} structures (\num{99493} atoms) as our training dataset and \num{822} structures (\num{25035} atoms) as our test dataset, respectively. 

For \gls{nep}, we set $r_{\rm c}^{\rm R}=6$ \AA{}, $r_{\rm c}^{\rm A}=4$ \AA{}, $N_{\rm neu}=50$, $N_{\rm gen}=5\times 10^5$, while keeping other hyperparameters as the defaults. 
For \gls{dp}, the \textsc{deepmd-kit} package (version 2.1.4) \cite{Wang2018cpc} is used, with the \verb"se_a" descriptor with a cutoff radius of 6 \AA. The dimensions of the embedding network are set to (25, 50, 100), and the fitting network dimensions are configured as (240, 240, 240). Initially, the weighting parameters for energy and forces are set to 0.02 and 1000, respectively, and are linearly adjusted to 1 for both during the training process. The training comprises $4\times 10^6$ steps, with a learning rate that is exponentially decreased from 10$^{-3}$ to 10$^{-8}$.
For \gls{mtp}, the \textsc{mlip} (version 2) package \cite{Novikov2021MLST} is used. The descriptor ``level'' for \gls{mtp} is set to 18, with a cutoff radius of 6 \AA. 
\autoref{table:mlps_compare} presents the performance metrics for the three \gls{mlp} models. 

\begin{table}[ht]
\centering
\setlength{\tabcolsep}{3Mm}
\caption{Comparison of the energy and force \glspl{rmse} and computational speed for \gls{mtp}, \gls{dp} (after model compression), and \gls{nep}. The computational speed is assessed by running \gls{md} simulations for $10^5$ steps in the $NVT$ ensemble for a graphene system containing \num{24800} atoms, using \textsc{gpumd} (\gls{nep}) or \textsc{lammps} \cite{Thompson2022cpc} with version 23 Jun 2022 (\gls{mtp} and \gls{dp}). 
For GPU-based tests (\gls{dp} and \gls{nep}), a single Nvidia RTX 3090 is used; for CPU-based tests (\gls{mtp}), 64 AMD EPYC 7H12 cores are used.}
\label{table:mlps_compare}
\begin{tabular}{llll}
\hline
\hline
Model              &  \gls{mtp}  & \gls{dp}     & \gls{nep}    \\
\hline
Energy-train (meV/atom) & 2.4   & 1.4   & 1.8   \\
Energy-test (meV/atom)  & 2.2   & 1.4   & 1.9  \\
Force-train (meV/\AA)  & 119   & 75    & 91 \\
Force-test (meV/\AA)   & 116   & 78    & 89 \\
Speed ($10^5$ atom-step/s)& 3.9 & 1.8  &  100  \\
\hline
\hline
\end{tabular}
\end{table}

We then conduct \gls{nemd} simulations to validate the implementations of heat current in the three \glspl{mlp} by checking the consistency between the accumulated heat in atoms within the transport region [cf. Eq.~(\ref{equation:j_pot_pair_stress})] and that obtained from the thermostats [cf. Eq.~(\ref{equation:Q_thermostat})]. The \gls{nemd} simulation procedure is similar to that as described in Sec.~\ref{section:nemd-si} for silicon. The transport is set to be along the armchair direction of a graphene sample with a width of 2.5 nm and a length of 426 nm (excluding the thermostatted regions). The data presented in Fig.~\ref{figure:heatflux} are sampled during the last 1.5 ns of the \gls{nemd} simulations, during which a steady state is achieved.

\bibliography{refs}

\end{document}